\documentclass[12pt,dvipsnames]{article}
 
\usepackage{iftex}
\ifLuaTeX
  \usepackage{polyglossia} 
  \usepackage{fontspec} 
  \setmainlanguage{english}
  \usepackage[nosort]{cite}
\else 
  \usepackage[utf8]{inputenc}
  \usepackage[T1]{fontenc}
  \usepackage[nosort]{cite}
  \usepackage[english]{babel} 
\fi

\usepackage[top=2cm,bottom=2cm,outer=2.5cm,inner=2.5cm,marginparwidth=2.5cm]{geometry}
\usepackage{fvextra} 
\usepackage{authblk} 

\ifLuaTeX
  \usepackage[unicode]{hyperref}
  \usepackage{pdftexcmds} 
  \makeatletter
  \let\pdfstrcmp\pdf@strcmp
  \makeatother
\else
  \usepackage{hyperref}
\fi

\usepackage{amsmath}
\usepackage{amsfonts}
\usepackage{amssymb}
\usepackage{tensor} 
\usepackage{braket} 
\usepackage{empheq} 
\usepackage{romanbar} 
\usepackage{dsfont} 
\usepackage{stmaryrd} 

\usepackage{lmodern} 
\usepackage{caption}
\usepackage{subcaption} 
\usepackage{titlesec} 
\usepackage{fancyhdr} 
\usepackage{abstract} 
\usepackage[super]{nth}
\usepackage[colorinlistoftodos,prependcaption,textsize=tiny]{todonotes} 
\usepackage{tocloft} 

\usepackage{graphicx} 
\usepackage{multirow} 
\usepackage{array} 
\usepackage{tikz}
\usepackage[compat=1.1.0]{tikz-feynman} 
\usepackage{pgf}
\usepackage{pgfplots} 
\usepackage{dynkin-diagrams}
\usepackage{hhline}

\usepackage{sidecap}
\sidecaptionvpos{figure}{c} 

\usetikzlibrary{arrows,shapes,calc,intersections,fadings,positioning,shadows.blur,decorations.pathreplacing,external,backgrounds}
\usepackage{currfile} 

\pgfplotsset{compat=1.16}
\usetikzlibrary{arrows.meta}
\tikzset{>=Latex}

\definecolor{Red-Teal-13-1}{HTML}{8b0000}
\definecolor{Red-Teal-13-2}{HTML}{b61d39}
\definecolor{Red-Teal-13-3}{HTML}{d84765}
\definecolor{Red-Teal-13-4}{HTML}{ef738b}
\definecolor{Red-Teal-13-5}{HTML}{fea0ac}
\definecolor{Red-Teal-13-6}{HTML}{ffd1c9}
\definecolor{Red-Teal-13-7}{HTML}{ffffe0}
\definecolor{Red-Teal-13-8}{HTML}{c7f0ba}
\definecolor{Red-Teal-13-9}{HTML}{9edba4}
\definecolor{Red-Teal-13-10}{HTML}{7ac696}
\definecolor{Red-Teal-13-11}{HTML}{5aaf8c}
\definecolor{Red-Teal-13-12}{HTML}{399785}
\definecolor{Red-Teal-13-13}{HTML}{008080}

\immediate\write18{mkdir -p figures}     
\tikzexternaldisable
\newcommand*\setmyname{
  \expandafter\tikzsetfigurename\expandafter{\currfilebase-}%
}
\AtBeginEnvironment{tikzpicture}{\setmyname}

\hypersetup{
pdfstartview={FitH},
pdftitle={},
pdfauthor={N. Drukker, M. Trepanier},
pdfsubject={},
pdfcreator={},
pdfkeywords={},
colorlinks=true,
citecolor=MidnightBlue,
linkcolor=MidnightBlue,
urlcolor=MidnightBlue
}

\numberwithin{equation}{section}
\newcommand\frontmatter{%
  \clearpage
  \pagenumbering{roman}
}

\newcommand\mainmatter{%
  \clearpage
  \pagenumbering{arabic}
}

\usepackage{mathtools}

\DeclareMathOperator\arctanh{arctanh}

\DeclareMathOperator{\tr}{tr}
\DeclareMathOperator{\sign}{sign}
\newcommand{\vev}[1]{\left\langle #1 \right\rangle}

\newcommand{\norm}[1]{\left| #1 \right|}

\newcommand{\diff}{\mathrm{d}}
\newcommand{\fin}{\text{finite}}
\newcommand{\ho}[2][]{\mathcal{O}(#2^{#1})}

\DeclareMathOperator{\arccoth}{arccoth}

\def\bD {\mathbb{D}}

\def\bH {\mathbb{H}}

\def\bO {\mathbb{O}}

\def\bQ {\mathbb{Q}}
\def\bR {\mathbb{R}}

\def\aQ{{\mathsf{Q}}}
\def\aS{{\mathsf{S}}}

\def\cA{{\mathcal{A}}}
\def\cN{{\mathcal{N}}}

\def\cL{{\mathcal{L}}}
\def\cR{{\mathcal{R}}}
\def\cO{{\mathcal{O}}}

\def\eF{\mathbf{F}}
\def\ePi{\mathbf{\Pi}}
\def\eE{\mathbf{E}}

\newcommand{\osp}{\mathfrak{osp}}
\newcommand{\sof}{\mathfrak{so}}

\newcommand{\bea}{\begin{eqnarray}}
\newcommand{\eea}{\end{eqnarray}}
\newcommand{\beq}{\begin{equation}}
\newcommand{\eeq}{\end{equation}}
\newcommand{\bal}{\begin{equation}\begin{aligned}}
\newcommand{\eal}{\end{aligned} \end{equation}}

\title{
Ironing out the crease}
\author[1]{Nadav Drukker\thanks{\href{mailto:nadav.drukker@gmail.com}{nadav.drukker@gmail.com}}}
\author[2]{Maxime
  Tr\'epanier\thanks{\href{mailto:trepanier.maxime@gmail.com}{trepanier.maxime@gmail.com}}}
\affil[1]{\it Department of Mathematics, King's College London,\protect\\London, WC2R
2LS, United Kingdom}
\affil[2]{\protect\it Institute for Theoretical and Mathematical Physics (ITMP),\protect\\ Moscow 119991, Russia}
\date{}

\begin{document}

\frontmatter
\maketitle
\thispagestyle{empty}

\begin{abstract}
The crease is a surface operator folded by a finite angle along an infinite line. Several realisations 
of it in the 6d ${\cal N}=(2,0)$ theory are studied here. It plays a role similar to the generalised 
quark-antiquark potential, or the cusp anomalous dimension, in gauge theories. We identify a 
finite quantity that can be studied despite the conformal anomalies ubiquitous with surface 
operators and evaluate it in free field theory and in the holographic dual. We also find a subtle 
difference between the infinite crease and its conformal transform to a compact observable 
comprised of two glued hemispheres, 
reminiscent of the circular Wilson loop. We prove by a novel application of 
defect CFT techniques for the $SO(2,1)$ symmetry along the fold 
that the near-BPS behaviour of the 
crease is determined as the derivative of the compact observable with respect to its angle, 
as in the bremsstrahlung function. We also comment about the lightlike limit of the crease in 
Minkowski space.
\end{abstract}

\mainmatter

\tableofcontents

\section{Introduction and summary}
\label{sec:intro}

In order to make progress on the 6d $\cN=(2,0)$ theory we need to either find avatars of the theory 
with lagrangian descriptions or use whatever limited tools there are in the absence of a lagrangian 
to study the theory. For examples of recent attempts in the former approach, see e.g. 
\cite{Sen:2019qit, Lambert:2020zdc, Mezincescu:2022hnb}. 
Here instead we follow the latter approach, extending our programme 
\cite{Drukker:2020dcz, Drukker:2020swu, Drukker:2020atp, Drukker:2020bes, Drukker:2021vyx} 
focused on the most natural observables on the theory---the 2d surface operators.

The simplest possible surfaces are the plane and sphere \cite{berenstein:1998ij} 
which can preserve the maximal amount 
of residual supersymmetry, 16 supercharges, as well as conformal symmetry and overall the 
algebra $\osp(4^*|2)^2$. Those are very natural starting points for exploring general surface 
operators, but contain little information on their own. The sphere suffers from a 
conformal anomaly so its expectation value depends on its radius $R$
\beq
\label{spherevev}
\vev{V_{S^2}}\propto R^{4c-2a_1}\,,
\eeq
where $a_1$ and $c$ are the anomaly coefficients given below in \eqref{anomalycoef}.

One can view the plane as the worldsurface of an infinite straight string. A natural question is 
to evaluate the potential between two strings, or the plane-antiplane configuration, as was studied already 
in \cite{maldacena:1998im}. Amongst other things, in this paper we enrich this configuration by 
allowing each plane to associate to a different scalar field, which gives a continuous parameter 
$\theta$ interpolating between the anti-parallel planes with $\theta=0$ and the BPS parallel planes 
at $\theta=\pi$.

\begin{figure}[tb]
  \centering
  \begin{subfigure}[t]{0.49\textwidth}
    \centering
    \includegraphics{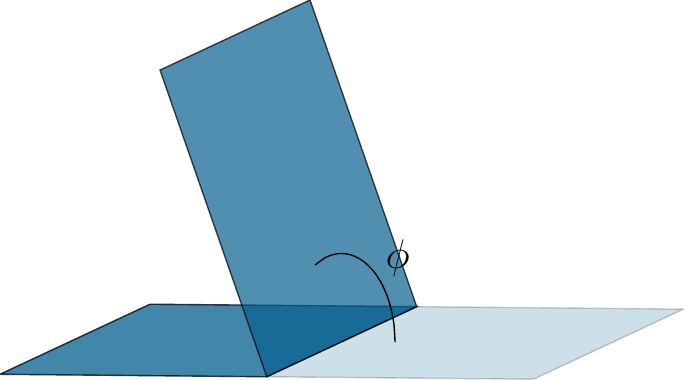}
    \caption{The infinite crease}
  \end{subfigure}
  \begin{subfigure}[t]{0.49\textwidth}
    \centering
    \includegraphics{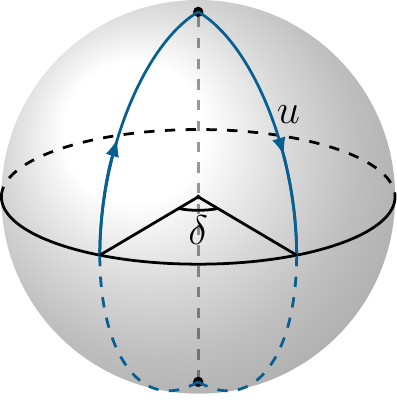}
    \caption{The compact crease}
  \end{subfigure}
  \caption[The compact crease]
  {On the left, the infinite crease with $\phi = 2\pi/3$. On the right,
  a representation of the same crease after a conformal transformation. Each of the half 
  planes is mapped to a hemisphere, here represented by the two arcs on $S^2$ (since we 
  cannot draw $S^3$). Each point along the arc represents a circle, which shrinks at the equator 
  ($z=0$ in the image) and is of maximal radius 1 at the north pole, where it is also glued to the 
  second hemisphere.}
  \label{fig:crease}
\end{figure}

A further generalisation of the system is the crease: two half-planes joined at an angle $\phi$ 
along a line (see figure~\ref{fig:crease}). When $\phi=0$ this is a single 1/2 BPS plane and when $\phi\to\pi$ it approaches 
the antiparallel planes. 

This is a clear analogue of the cusped Wilson loop or ``generalised quark-antiquark'' system 
studied in \cite{drukker:1999zq, Drukker:2011za, correa:2012at, Drukker:2012de, Correa:2012hh, Gromov:2012eu}. 
Another natural 2d analogue of the cusp is a cone. 
Under a conformal transformation to $S^5\times\bR$ the cone becomes a cylinder which is a 
circle in $S^5$ extending along the $\bR$ direction. BPS versions of such operators 
were studied in \cite{mezei:2018url} and the limit of a BPS cylinder in $\bR^6$ in \cite{Drukker:2021vyx}. 

The crease configuration with the additional angle $\theta$ and several related 
surfaces are the subject of this paper. Our goal is to define and evaluate a meaningful 
physical quantity $U(\phi,\theta)$, analogous to the generalised quark-antiquark potential 
or cusp anomalous dimension.
This quantity can be viewed as the potential between the two half planes, but on a more 
abstract level, it is a finite observable of the $\cN = (2,0)$ theory that can be (at least partially) 
evaluated. As we now have a good handle on the anomaly coefficients, we are behoven 
to find more refined quantities associated to surface operators and this is exactly 
such an observable.

In the remainder of the introduction we define the quantity $U(\phi,\theta)$ and discuss its 
salient features. We then proceed to evaluate it in the subsequent sections.

The observable we study has several realisations. First, as an infinite flat crease between two half-planes 
parametrised by $\tau, \sigma\in\bR$
\beq
\label{infinitecrease}
x^\mu=\begin{cases}(\tau,\sigma,0,0,0,0)\,,&\sigma\leq0\,,\\
(\tau,\sigma\cos\phi,\sigma\sin\phi,0,0,0)\,,&\sigma\geq0\,.\end{cases}
\eeq
The surface also breaks the R-symmetry. It is natural 
to have along each half-plane a uniform breaking, expressed by a unit 5-vector $n^I$ as
\beq
n^I =
  \begin{cases}
    (1,0,0,0,0) \,,& \sigma \leq 0\,,\\
    (\cos\theta,\sin\theta,0,0,0) \,,& \sigma \geq 0\,.
  \end{cases}
\label{eqn:sscalarparam}
\eeq
The 1/2 BPS plane is clearly at $\theta=\phi=0$. If $\theta=\pm\phi\neq0$ this operator 
is 1/4 BPS and according to the classification of BPS surface operators in \cite{Drukker:2020bes} 
it is of Type-$\bR$, since it has translation invariance (it is also of Type-$\bH$, which is 
utilised in Appendix~\ref{app:calibration}). 
For $\theta\neq\pm\phi$ no supersymmetry is preserved, but still in all 
cases the surface operator preserves the 1d conformal group $SO(2,1)$, the transverse 
rotation group $SO(3)$ as well as $SO(3)$ R-symmetry. We refer to a surface operator with 
this geometry as an \emph{infinite crease}.

Under a conformal transformation a plane is mapped to a sphere and a half-plane to a 
hemisphere. The crease is mapped into a pair of hemispheres inside $S^3$ joined along 
a large circle. Explicitly we can express $S^3\subset\bR^4$ as
\bal
  x^1 &= R \cos u \cos v \,, \qquad
  x^2 = R \cos u \sin v\,, \qquad\\
  x^3 &= R \sin u  \cos w\,, \qquad
  x^4 = R \sin u \sin w\,,
  \label{eqn:screaseparam}
\eal
with $u\in[0,\pi/2]$ and $v,w\in[0,2\pi)$. Each hemisphere is specified by a fixed
$w$, which we take to be $w=\phi/2$ and $w=\pi-\phi/2$.  They are clearly
joined at an angle $\pi-\phi$ along the $u=0$ circle. In the following we
refer to this as the \emph{spherical crease}.

Again we can associate to each of the hemispheres a different scalar coupling
separated by an angle $\theta$. 
A simple way to write it is by extending the domain of $u$ to $[-\pi/2,\pi/2]$ with $x^4=R
\sin|u| \sin(\phi/2)$ and
\begin{align}
  n^1 = \cos\frac{\theta}{2}\,, \qquad
  n^2 = \sign{u} \sin\frac{\theta}{2}\,.
  \label{eqn:screasescalarparam}
\end{align}
As with the infinite crease, when $\theta = \phi$ this is 1/4 BPS, but now it is not invariant under 
Poincar\'e supercharges but only combinations with special conformal 
transformations, making it of Type-S, in the classification of \cite{Drukker:2020bes}.

Translation invariance of the infinite crease suggests that any observable we calculate 
would be extensive in its length, but in fact any calculation preserving the conformal symmetry 
should be extensive in the area of the half-planes when endowed with a hyperbolic metric. 
It is thus natural to conformally transform $\bR^6$ to $AdS_2\times S^4$ as%
\footnote{By $AdS_2$ we always mean Euclidean $AdS_2$, i.e. $\bH_2$.}
\beq
\sum_{\mu=1}^6 \diff x^\mu\diff x^\mu
=r^2\left(\frac{\diff x^1\diff x^1+\diff r\diff r}{r^2}+\diff\Omega_4^2\right)\,,
\qquad
r^2=\sum_{\mu=2}^6 (x^\mu)^2\,.
\label{PAdS2S4}
\eeq
The crease \eqref{infinitecrease} is the union of two copies of $AdS_2$ at points on 
$S^4$ separated by the angle $\pi - \phi$. 
In this picture, the crease singularity is replaced by the jump at the 
boundary of the hyperbolic spaces.

Likewise for the spherical crease, if we start with the theory on $\bR^6$, stereographically 
project to $S^6\subset\bR^7$ given by \eqref{eqn:screaseparam} with $R=1$ and the $w$ 
circle extended to $S^4$, we can rewrite the $S^6$ metric as
\beq
\diff u^2 + \cos^2u\, \diff v^2 + \sin^2u\, \diff \Omega^2_4
= \frac{1}{\cosh^2\rho}
\left( \diff \rho^2 + \sinh^2\rho\, \diff v^2 + \diff \Omega^2_4 \right).
\label{eqn:conftrs6ads2}
\eeq
This is accomplished with $\cosh\rho=1/\sin u$.

The area of $AdS_2$ diverges so any quantity we evaluate should also suffer from those 
divergences. The natural regularisation of the area gives zero for the upper half 
plane model in \eqref{PAdS2S4} and $-2\pi$ for the disc model in \eqref{eqn:conftrs6ads2}. 

In the latter case we can therefore unambiguously define the potential $U(\phi,\theta)$ to be the 
density that arises in the calculation, such that when using the disc model of $AdS_2$, 
the regularised expectation value of the spherical crease is $\exp[2\pi U(\phi,\theta)]$. In the case 
of the infinite crease, where the regularised area vanishes, we can still try to define 
the potential via the density, but it is more subtle, as it is multiplied by zero in the 
final expression.

There is one more subtlety though, which is the conformal anomaly arising in calculating 
surface operators. The expectation value of a general surface operator suffers from logarithmic 
divergences proportional to the integral of the anomaly density
\cite{Schwimmer:2008yh, Drukker:2020dcz}
\beq
\label{anomalydens}
\cA=\frac{1}{4\pi}\left(a_1\cR
+a_2(H^2+4\tr P)+b\tr W+c(\partial n)^2
\right),
\eeq
where $\cR$ is the Ricci scalar of the induced metric $h_{ab}$ on the surface, $H^\mu$ the mean curvature 
vector, $P=h^{ab}P_{ab}$ is the trace of the pullback of the Schouten tensor, $W$ is the pullback of the 
Weyl tensor and $\partial n$ is the gradient of the scalar coupling $n^I$. The prefactors are known as 
anomaly coefficients and are intrinsic properties of the theory and type of surface operator. 
It is known~\cite{Drukker:2020atp} that for this theory $b=0$ and $c=-a_2$. In
the abelian theory the coefficients are~\cite{Drukker:2020dcz}
\beq
a_1^{(1)}=\frac{1}{2}\,,\qquad
c^{(1)}=\frac{1}{2}\,,
\label{eqn:anomalycoeffu1}
\eeq
and for a surface operator in the fundamental representation in the nonabelian theory with algebra 
$A_{N-1}$ they are~\cite{graham:1999pm,Estes:2018tnu,Chalabi:2020iie,Wang:2020xkc}
\beq
\label{anomalycoef}
a_1^{(N)}=\frac{1}{2}-\frac{1}{2N}\,,\qquad
c^{(N)}=N-\frac{1}{2}-\frac{1}{2N}\,.
\eeq
The classical holographic calculation \cite{graham:1999pm} just captures $a_1\sim0$ and $c\sim N$.

One may be concerned that all the creases that we study have an anomaly arising from the curvature 
singularity at the crease itself where $H$ and $\partial n$ have delta-function contributions. 
The formula \eqref{anomalydens} above does not apply for singular cases, 
and as we confirm in Section~\ref{sec:creaseanomaly}, indeed the singularity does not lead to an anomaly. 
Thus we see that the infinite crease has no anomaly and we can therefore try to define 
$U(\phi,\theta)$ as the potential density, but there is still the issue that its regularised 
integral vanishes. The spherical crease is anomalous, but since the 
anomaly is local and does not get a contribution from the crease singularity, it is independent of 
the angles, and the same as for 
the sphere with $\phi=\theta=0$ and equal to $2a_1-4c$ \eqref{spherevev} We then can define
\beq
U(\phi,\theta)=\frac{1}{2\pi}\big(  \log{\vev{V_{\phi,\theta}}} -
\log{\vev{V_{S^2}}}\big)\,,
  \label{eqn:potentialdefn}
\eeq
with the $2\pi$ factor arises from the area of the Poincar\'e disc.

It is most obvious that the anomaly does not depend on $\phi$ and $\theta$ in the $AdS_2\times S^4$ picture. 
Here there are two disconnected $AdS_2$ factors with vanishing extrinsic curvature, so by the Gauss-Codazzi 
equation, the second fundamental form is twice the Ricci scalar, which is a constant $-2$, so the anomaly 
density is $(a_1+2a_2)\cR/4\pi$ which after integration (and with the appropriate boundary term) 
gives for each $AdS_2$ either zero or $a_1-2c$, depending on the topology.

Having defined $U(\phi,\theta)$ we proceed in the rest of the paper to evaluate
it: First 
in the free abelian version of the $\cN = (2,0)$ theory in Section~\ref{sec:free} 
and then in the holographic dual in terms of M2-branes in $AdS_7\times S^4$ in 
Section~\ref{sec:AdS}. The results of the two calculations are not identical, so 
this observable is an interesting quantity that depends on the rank of the algebra 
underlying the theory 
(as well as a representation of that algebra to which the surface operator is related) 
and $\phi$ and $\theta$.

If we focus on the BPS case, when $\phi=\theta$ the expressions do not vanish, as 
one may have expected. Instead we find
\beq
U(\phi,\phi) = \frac{C}{\pi} \log{\cos\frac{\phi}{2}}\,.
\label{eqn:potentialBPS}
\eeq
In both the free theory and at leading 
order at large $N$, the constant $C$ is the same as the anomaly coefficient $c$ in \eqref{anomalycoef}. 
Interestingly, we find the same functional dependence on $\phi$ in both calculations.

Like in the recently studied cases of the torus and cylinder \cite{Drukker:2021vyx} or 
those in \cite{mezei:2018url}, this is a finite nonzero quantity associated to a BPS observable, 
similar in that regard to the circular Wilson loop in $\cN=4$ SYM in 4d 
\cite{erickson:2000af, drukker:2000rr, pestun:2007rz}.

This can be extended to the near-BPS limit, where
\beq
U(\phi,\theta) = \frac{C}{\pi} \log{\cos\frac{\phi}{2}}
-\frac{C}{2\pi}(\phi-\theta)\tan\frac{\phi}{2}+\ho[2]{(\theta-\phi)}\,.
\label{eqn:potentialnearBPS}
\eeq
For small $\phi$ and $\theta$, we can show that there is no $\phi\theta$ term 
and find
\beq
U(\phi,\theta) = -\frac{C}{8\pi}(2\phi^2-\theta^2) + \dots
\label{eqn:potentialnearplane}
\eeq
This expression is similar in spirit to the bremsstrahlung formula of 
\cite{correa:2012at, Fiol:2012sg}, which arises from the nearly straight cusped 
Wilson loop. The expression \eqref{eqn:potentialnearBPS}, which can 
also be written as
\beq
U(\phi,\theta)=\left(1+(\phi-\theta)\frac{\diff}{\diff\phi}\right)U(\phi,\phi)+\cO((\phi-\theta)^2)\,,
\label{eqn:potentialnearBPS2}
\eeq
 is then akin to the 
``generalised bremsstrahlung formula'' valid for near-BPS cusps in $\cN=4$ SYM. 
In Section~\ref{sec:dCFT} we reproduce these results from a defect CFT analysis 
and show that indeed $C=c$, the anomaly coefficient.

One more special example of the crease, when the two half-planes approach lightlike 
surfaces, is studied in Section~\ref{sec:lightlike}. This is very similar to the case of 
the lightlike cusp~\cite{Kruczenski:2002fb} as well as its extension to the 4-cusp solution of 
Alday and Maldacena~\cite{Alday:2007hr}. In our case the analogue of the 4-cusp solution 
involves 2 lightcones joined at a circular crease. We discuss some properties of those solutions 
and evaluate their action.

We comment about possible applications and extensions of this work in Section~\ref{sec:conclusion}.

\section{Free theory}
\label{sec:free}

\subsection{The spherical crease}

We start with the abelian version of the $\cN=(2,0)$ theory
\cite{Witten:1995em,Kaplan:1995cp}. 
The field content of the theory is
\begin{itemize}
\item
$B_{\mu\nu}$, a two-form with self-dual field strength.
\item
$\Phi^I$, five scalar fields.
\item
$\Psi$, eight symplectic Majorana fermions.
\end{itemize}

The surface operators of this theory are constructed explicitly
in~\cite{Drukker:2020dcz,gustavsson:2004gj}. They are built out of the fields $B$ and $\Phi$
and are written as
\beq
V_{\phi,\theta} =
\exp\int_\Sigma \sqrt{h}\left[\frac{i}{2} B_{\mu\nu} \partial_a x^\mu \partial_b x^\nu\varepsilon^{ab}
- n^I \Phi^I \right] \diff^2 \sigma\,.
\label{eqn:defnV}
\eeq
Here $a,b$ are the indices of the coordinates $\sigma^a$ parametrising the
surface; in our definition of the crease~\eqref{eqn:screaseparam},
we take $\sigma^a=(u,v)$. $h_{ab}$ is the induced metric, $\varepsilon^{ab}$ is
the antisymmetric tensor normalised with a factor of $1/\sqrt{h}$ and $n^I$ are
the scalar couplings, which in our examples are given
in~\eqref{eqn:screasescalarparam}. 

Despite the subtleties related to the self-duality of $B$, one can derive propagators for the 
bosonic fields in flat space. Including a short distance regulator $\epsilon$ they are 
\cite{henningson:1999xi,gus03,gustavsson:2004gj,Drukker:2020dcz}
\begin{subequations}
\begin{align}
  \label{Propagator_Phi}
  \vev{\Phi^I(x) \Phi^J(y)} &= \frac{\delta^{IJ}}{2\pi^2 (\norm{x-y}^2+2\epsilon^2)^2}\,, \\
  \vev{B_{\mu\nu}(x)B_{\rho\sigma}(y)} &=
  \frac{\delta_{\mu\rho}\delta_{\nu\sigma}-\delta_{\mu\sigma}\delta_{\nu\rho}}{2\pi^2( \norm{x-y}^2+2\epsilon^2)^2}\,.
\label{eqn:CurvedSpacePropagators}
\end{align}
\end{subequations}

We can use these propagators to calculate the expectation value of the spherical
crease. It is easy to show that because the theory is free, the expectation value
is given by (the exponential of) the propagator integrated over 2 copies of the
surface. There are two contributions, coming from integrating the propagator
over two copies of the same hemisphere and the opposites hemispheres.
Using the explicit parametrisation~\eqref{eqn:screaseparam},
\eqref{eqn:screasescalarparam}, the log of the expectation value can be expressed as
\beq
\log{\vev{V_{\phi,\theta}}} =
\frac{1}{2}\int R^4 \cos{u}\cos u'\, \diff u\, \diff v\, \diff u'\, \diff v'
\left(2\Delta_{\pi,\pi}(u, v; u',v')
+2\Delta_{\phi,\theta}(u, v; u',v')\right)\,,
\eeq
where the combined propagator is
\beq
\label{Delta}
\Delta_{\phi,\theta}(u, v; u',v')= \frac{1}{8\pi^2}
\frac{\cos{\theta} -\cos u \cos{u'} \cos(v-v')\cos{\phi} +\sin{u} \sin{u'}}
{R^4\left(1-\cos{u} \cos{u'} \cos{(v-v')}+\sin{u} \sin{u'}\cos{\phi}
+ \epsilon^2/R^2\right)^2}\,.
\eeq

We can perform the $v$ and $v'$ integral using equation (2.554)
of~\cite{gradshteyn2014table}
\beq
\int_0^{2\pi} \diff v\,\diff v' \frac{A + B\cos(v-v')}{(a + b \cos(v-v'))^2}
=2\pi\int_0^{2\pi} \diff v \frac{A + B\cos{v}}{(a + b \cos{v})^2}
=4\pi^2 \frac{a A - b B}{\left[a^2-b^2\right]^{3/2}}\,, \qquad a>b\,.
\eeq
With the appropriate $A$, $B$, $a$ and $b$ and replacing $s=\sin u$ and $s'=\sin u'$ 
this is (with $R=1$)
\bal
\label{partialDint}
\int\diff v\,\diff v'\, \Delta_{\phi,\theta}
&=\frac{1}{2}
\frac{(\cos{\theta}-\cos{\phi})(1+\epsilon^2-ss'\cos{\phi} )}
{\left[(1+\epsilon^2+ss'\sin\phi)^2-(1-s^2)(1-s^{\prime2})\right]^{3/2}}
\\&\quad
+
\frac{1}{2}
\frac{ \cos{\phi}( s^2+s^{\prime2}+\epsilon^2) +ss' (1+\epsilon^2+\cos^2{\phi})}
{\left[(1+\epsilon^2+ss'\sin\phi)^2-(1-s^2)(1-s^{\prime2})\right]^{3/2}}\,.
\eal
Integrating this gives
\beq
\label{Dintegral}
\int \diff s\,\diff v\,\diff s'\,\diff v'\,\Delta_{\phi,\theta}
=( \cos{\theta} - \cos{\phi}) \left[ \frac{R \phi}{2^{3/2}\,\epsilon \sin{\phi}}
-\frac{1}{4 \cos^2(\phi/2)}
\right] + \log{\left(2 \cos\frac{\phi}{2}\right)}+\ho{\epsilon}\,.
\eeq
The contribution from two copies of the same hemisphere, for which
$\phi=\theta=\pi$,
has further UV divergences and we cannot rely on the regularised 
result of the integral above. Instead, after substracting $\epsilon^2$ from the numerator in 
\eqref{Delta} we find that \eqref{partialDint} is now
\beq
\int\diff v\,\diff v' \Delta_{\pi,\pi}
=\frac{1}{2 \sqrt{\epsilon^4+2 \epsilon^2 (1-s s')+(s-s')^2}}\,,
\eeq
and
\beq
\int \diff s\,\diff v\,\diff s'\,\diff v'\,\Delta_{\pi,\pi}=-\log\frac{\sqrt2\,\epsilon}{R}+\ho{\epsilon}\,.
\eeq
Combining with \eqref{Dintegral} and removing the linear divergence we find
\beq
\label{VfreeVEV}
\log{\vev{V_{\phi,\theta}}} =
\log{\frac{\sqrt2\, R \cos(\phi/2)}{\epsilon}} +
\frac{\cos\phi - \cos\theta}{4\cos^2(\phi/2)}
+ \ho{\epsilon}\,.
\eeq
Setting $\phi = \theta = 0$ indeed reproduces the result for the
sphere~\eqref{spherevev} with the anomaly
coefficients~\eqref{eqn:anomalycoeffu1}.

We can then take the difference to the regular sphere case to read off the
potential~\eqref{eqn:potentialdefn}
\beq
U^{(1)}(\phi,\theta) =
\frac{1}{2\pi} \log{\cos\frac{\phi}{2}} + \frac{\cos\phi - \cos\theta}{8\pi
  \cos^2(\phi/2)}\,.
\label{eqn:potentialabelian}
\eeq

\subsection{The crease singularity}
\label{sec:creaseanomaly}

In the above derivation we found a linear $R/\epsilon$ divergence. More
confusing is the relation of this result to the anomaly formula
\eqref{anomalydens}. Along either the infinite or the compact crease, the mean
curvature vector $H$ has a delta function singularity, invalidating the
formula. One may worry that under regularisation we would get a term similar to
$\epsilon^{-1}\log\epsilon$, which we did not find above. We examine this
apparent contradiction here.

For simplicity we choose to analyse the case of the sphere $\phi=0$ with
$\theta\neq0$. $n^I$ is given in~\eqref{eqn:screasescalarparam} and
clearly is singular at $u=0$; we can regularise it by replacing the 
step function $\sign(u)$ by a continuous function $f_l(u)$ with a regulator $l$. 
We can take for instance
\beq
f_l(u) =
\begin{cases}
u/l\,, & -l\leq u\leq l\,,\\
\sign(u)\,, & \text{otherwise}\,.
\end{cases}
\eeq
As $l \to 0$ this reduces to the step function.

It is easy to check that now $|\partial n|\sim 1/l$ for $|u|\leq l$ and the anomaly formula indeed gives 
a $l^{-1} \log{\epsilon}$ divergence. Any other regulator $f_l$ would give the same behaviour 
up to a numerical factor.

The mechanism by which this discrepancy is resolved is easy to understand in
this example.  Using \eqref{Delta}, the expectation value for the
regularised crease in the free theory is
\beq
  \frac{1}{16 \pi^2} \int \cos{u}\cos u' \,\diff u\, \diff v\, \diff u'\, \diff v'
  \frac{n^I(u) n^I(u') - \cos{u} \cos{u'} \cos{(v-v')} - \sin{u} \sin{u'}}
  {(1-\cos{u}\cos{u'}\cos{(v-v')}-\sin{u} \sin{u'} + \epsilon^2/R^2)^2}\,.
\eeq
Instead of evaluating the exact expression, we expand in a power series in $\theta$. 
For $\theta = 0$ this is the regular sphere, so there is no singularity, only the usual $-\log\epsilon$ 
anomaly. The first nontrivial term is of order $\theta^2$ and it is enough 
to capture the $l^{-1} \log{\epsilon}$ divergence. After expanding, the $v,v'$ integrals
are performed as before and substituting again $s = \sin{u}$ and $s' = \sin{u'}$ 
we are left with
\beq
-\frac{1}{32}\int_{-1}^{1} \diff s\, \diff s'\,
\frac{(1 - ss') (f_l(s)-f_l(s'))^2}
{\left( (s-s')^2 + (1-s s') \epsilon^2 + \epsilon^4
\right)^{3/2}}\,.
\eeq
This integral can be performed and expanding the result first in $\epsilon$ with fixed $l$
and then expanding in small $l$ gives
\beq
\frac{1}{16l} + \frac{1}{8l} \log\frac{\epsilon}{4l} +\fin.
\eeq
For fixed $l$ and $\epsilon\to0$ we indeed find the result $l^{-1}\log\epsilon$ 
predicted by the anomaly formula. The $1/l$ and $l^{-1}\log l$ terms are then regarded as 
finite contributions.

But this expression does not lead to a $\epsilon^{-1}\log\epsilon$ divergence, as when we take 
$l\to \epsilon$, the $\log l$ compensates for the $\log\epsilon$ and all we are 
left with is a linear divergence $1/l\to1/\epsilon$, in agreement with the explicit calculations. 
Furthermore, there are no new $\log\epsilon$ terms for $l\sim\epsilon$, and all the anomaly  
for the sphere with $\theta\neq0$ is the same as the usual sphere, as indeed we found in 
\eqref{VfreeVEV}.

Of course in this example we only look at the term of order $\theta^2$ for a
simple choice of $f_l$. We did not calculate the regularised crease for any
$\phi$ and $\theta$ but we expect the same mechanism to reconcile the anomaly formula
prediction with the results for the singular crease. Indeed $\epsilon$ in the
logarithm must appear with some other scale, and the scale relevant for the
singularity is $l$.

The observation that the crease does not give additional $\log^2{\epsilon}$ or
$\log{\epsilon}$ divergences was also made in the context of holographic
entanglement entropy in~\cite{myers:2012vs,Bueno:2019mex,klebanov:2012yf}.

\subsection{The infinite crease}
\label{sec:pert-inf}

We defined the potential \eqref{eqn:potentialdefn} in terms of the spherical crease and in all our calculations 
it is identical also to its $AdS_2$ realisation. Here we examine what can be calculated for the infinite crease.

Repeating the calculation with this geometry, the interesting contribution comes 
from propagators between different half-planes in \eqref{infinitecrease}. Using 
\eqref{Propagator_Phi}, \eqref{eqn:CurvedSpacePropagators}
it reads
\begin{align}
\label{inf-int}
  \log{\vev{V_{\phi,\theta}^\mathrm{inf}}}
  = \frac{\cos{\theta}-\cos{\phi}}{2 \pi^2} \int_{-\infty}^\infty \diff \tau \,\diff \tau'\int_{0}^\infty 
  \frac{\diff \sigma\,\diff \sigma' }{[(\tau - \tau')^2 +
  (\sigma-\sigma')^2+4\sigma\sigma'\cos^2(\phi/2)+2\epsilon^2]^2}\,.
\end{align}
Performing the integrals over $\tau'$, $\sigma'$ and $\sigma$, we obtain
\begin{align}
\label{inf-answer}
\log\vev{V_{\phi,\theta}^\text{inf}}
= \frac{\cos\theta - \cos\phi}{2^{3/2} \pi \sin\phi}\phi 
\int_{-\infty}^\infty\frac{\diff\tau}{\epsilon}\,.
\end{align}
The resulting integral gives the overall length $L/\epsilon$, which as a linear divergence can be removed. 
But we wish to view it instead as the volume of $AdS_2$ in the Poincar\'e patch
\beq
\label{poincare}
\int_{-\infty}^\infty\frac{\diff\tau}{\epsilon}
=\int_{-\infty}^\infty\diff\tau\int_\epsilon^\infty\frac{\diff \sigma}{\sigma^2}\,.
\eeq
Then we view the prefactor in \eqref{inf-answer} as representing the potential between the two planes
\beq
U^\text{inf}_{\phi,\theta} = \frac{\cos\theta - \cos\phi}{2^{3/2} \pi \sin\phi}\phi\,,
\eeq
where we did not divide by $1/2\pi$ as in \eqref{eqn:potentialdefn}, since here we divide by the 
divergent area of the upper half plane instead of the finite regularised area of the Poincar\'e disc.

Interestingly, the answer is different from the case of the spherical crease \eqref{eqn:potentialabelian}, 
and here it vanishes in the BPS case $\phi=\theta$. Also, it has the same functional form as the 
cusped Wilson loop in $\cN=4$ SYM at one-loop order \cite{Drukker:2011za}.

A naive integration of \eqref{inf-int} without the $\epsilon$ regulator is clearly divergent. But for 
fixed $\sigma$ and $\tau$ one can perform the full $\sigma'$, $\tau'$ integral to find
\beq
\frac{\cos\theta - \cos\phi}{8 \pi \cos^2(\phi/2)}\,,
\eeq
which is now the same as \eqref{eqn:potentialabelian} except for the $1/(2\pi)\log\cos(\phi/2)$ term. 
The remaining $\tau$, $\sigma$ integrals are as in \eqref{poincare}.

These two calculations are consistent with using two different cutoffs $\epsilon$ and $\epsilon'$ related by
\beq
\epsilon=\frac{2^{3/2}\cos^2(\phi/2)\phi}{\sin\phi}\epsilon'\,.
\eeq
Note that to reproduce the missing term from \eqref{eqn:potentialabelian}, we would need to include a 
$\theta$ dependence in cutoff, which is not very natural as a field theory cutoff. So we conclude that the result 
is rather ambiguous and scheme dependent, but in any case vanishes in the BPS case. 

We look at the holographic dual of the infinite crease in Appendix~\ref{app:poincare} and find similar ambiguities.

\section{Holography}
\label{sec:AdS}

\subsection{The setup}
\label{sec:setup}

The spherical crease is inside $S^3$ and involves two scalar fields. It can therefore 
be described by an M2-brane inside $AdS_4\times S^1\subset AdS_7\times S^4$. 
Writing the $AdS_4$ factor as an $AdS_2\times S^1$ foliation, this is
\beq
\label{ads-metric1}
\diff s^2 = 4L^2 \left[
\diff \nu^2 + \cosh^2{\nu}( \diff \rho^2 + \sinh^2 \rho\, \diff v^2)
+ \sinh^2 \nu\,\diff \varphi^2
\right]
+ L^2 \diff \vartheta^2 \,.
\eeq

Defining $\sinh\nu=1/t$ the metric becomes
\beq
\label{metric-t}
\diff s^2 = \frac{4L^2}{t^2}\left[
\frac{\diff t^2}{1+t^2} + (1+t^2)( \diff \rho^2 +\sinh^2 \rho\, \diff v^2)
+ \diff \varphi^2
\right]
+ L^2 \diff \vartheta^2 \,.
\eeq
The M2-brane can be parametrised by the coordinates of $AdS_2$: $\rho,v$, and 
in addition $t$. Because the crease has $AdS_2$ symmetry, we look for brane
embeddings depending only on $t$, so we should solve for $\varphi(t)$ and 
$\vartheta(t)$. This description is not single valued; $\varphi(t)$ and
$\vartheta(t)$ have two branches corresponding to each hemisphere of the
spherical crease. At the boundary of $AdS$ ($t\to0$), $\varphi$ and $\vartheta$
should approach their asymptotic values (shifted to be symmetric around $\pi/2$ and 0)
\beq
\varphi(0) = \phi/2\,,\ \pi-\phi/2\,, 
\qquad
\vartheta(0) = \pm\theta/2\,.
\label{eqn:creasebc1}
\eeq
A single branch of the solution has $t\in(0,t_\text{max}]$ and both branches
meet at $t_\text{max}$
(see for example Figure~\ref{fig:exampleholo})
\beq
\varphi(t_\text{max}) = \pi/2\,, \qquad
\vartheta(t_\text{max}) = 0\,.
\label{eqn:creasebc2}
\eeq

The action is
\beq
S_\text{M2} =
8 T_\text{M2} L^3 \int \sinh{\rho} \,\diff \rho\, \diff v\, \diff t\,
\frac{1+t^2}{t^3} \sqrt{\frac{1}{1+t^2}+ \varphi'^2 + \frac{t^2}{4}\vartheta'^2}\,.
\label{eqn:creaselagrangian}
\eeq
Here $T_\text{M2}$ is the M2-brane tension, which by the $AdS$/CFT dictionary~\cite{maldacena:1997re} 
is $N/4\pi L^3$.

This lagrangian has two conserved quantities: the canonical momenta 
conjugate to $\varphi$ and~$\vartheta$
\beq
J_\varphi=
\frac{(1+t^2)\varphi'}{t^3\sqrt{\frac{1}{1+t^2}+ \varphi'^2 + \frac{t^2}{4}\vartheta'^2}}\,, \qquad
J_\vartheta=
\frac{(1+t^2)\vartheta'}{4t\sqrt{\frac{1}{1+t^2}+ \varphi'^2 +\frac{t^2}{4}\vartheta'^2}}\,.
\label{eqn:creaseeom}
\eeq
Inverting these relations gives the first order equations of motion
\bal
\vartheta'(t) &= \frac{4 J_\vartheta t}{\sqrt{(1+t^2)(1+2t^2+(1-4 J_\vartheta^2)t^4-J_\varphi^2 t^6)}}\,,\\
\varphi'(t) &= \frac{J_\varphi t^3}{\sqrt{(1+t^2)(1+2t^2+(1-4 J_\vartheta^2)t^4-J_\varphi^2 t^6)}}\,.
\label{eqn:creaseeom2}
\eal

In order for the brane to be smooth, we need that at $t_\text{max}$ both $\varphi',
\vartheta' \to \infty$ together. This determines $t_\text{max}$ as one of the roots
of the polynomial in the denominator of~\eqref{eqn:creaseeom2}.  More precisely,
because both $J_\varphi$ and $J_\vartheta$ are real, it is simple to show that the polynomial only has a
single positive root. It can be factorised as
\beq
1+2t^2+(1-4 J_\vartheta^2)t^4 -J_\varphi^2 t^6
=-J_\varphi^2 (t^2-\tau_1)(t^2-\tau_2)(t^2-t_\text{max}^2)\,.
\label{eqn:creasepoly}
\eeq
The roots $\tau_1$, $\tau_2$ can be expressed explicitly in terms of $J_\varphi$, $J_\vartheta$ using the
cubic formula, but we leave them implicit. They are either complex or real and
negative. The root $t_\text{max}$ is real and positive.

\subsection{BPS case}
\label{sec:BPS}

As we show in Appendix~\ref{app:calibration}, in the BPS case, the set of equations
obtained in~\cite{Drukker:2020bes} relate the two momenta
\beq
\label{BPS}
2J_\vartheta=\pm J_\varphi\,,
\eeq
or equivalently $\vartheta' = \pm2 \varphi'/t^2$. On the first branch, we expect $\varphi$ to be monotonously increasing and
$\vartheta$ to be decreasing, 
and because their derivatives are proportional to the conserved momenta \eqref{eqn:creaseeom2}, 
we should take the solution with the negative sign.

In particular this means that the polynomial
appearing in the denominator of~\eqref{eqn:creaseeom2}
\beq
(1+t^2)\left[(1+t^2)^2-(4J_\vartheta^2+J_\varphi^2t^2)t^4\right],
\label{eqn:poly}
\eeq
has degenerate roots at $t^2 = -1$ and simplifies to
\beq
(1+t^2)^2\left(1+t^2-J_\varphi^2t^4\right).
\eeq
The factorisation of an additional factor of $1+t^2$ implies that in this case,
the equations of motion~\eqref{eqn:creaseeom2} contain the square root of a
quadratic polynomial, so their solution is trigonometric, rather than elliptic.

Comparing with~\eqref{eqn:creasepoly}, we find a simple expression for
$t_\text{max}$
\begin{align}
  t_\text{max}^2 = \frac{1 + \sqrt{1 + 4 J_\varphi^2}}{2 J_\varphi^2}\,.
  \label{eqn:umaxBPS}
\end{align}
Integrating $\vartheta'$ in~\eqref{eqn:creaseeom2} and imposing the boundary
conditions~\eqref{eqn:creasebc2} at $t_\text{max}$, we obtain
\begin{align}
  \vartheta = \arctan \left( \frac{2 J_\varphi \sqrt{1 + t^2 - J_\varphi^2 t^4}}
  {1 + (1+2J_\varphi^2) t^2} \right)\,,
  \label{eqn:bpssoltheta}
\end{align}
and similarly for $\varphi$
\begin{align}
  \varphi = \vartheta + \arcsin \left( \frac{J_\varphi t^2}{\sqrt{1 + t^2}}
  \right)\,.
  \label{eqn:bpssolphi}
\end{align}

We can read the value of $\phi$, $\theta$ from the solution by setting $t=0$:
\eqref{eqn:bpssolphi} gives $\phi=\theta$ as expected (the case
$\phi = -\theta$ is obtained by picking the positive sign in~\eqref{BPS}
instead), and \eqref{eqn:bpssoltheta} expresses $J_\varphi$
in term of the angle $\phi$
\begin{align}
  J_\varphi = \frac{1}{2} \tan\frac{\phi}{2}\,.
\end{align}

Finally, with these solutions in hand we can evaluate the
action~\eqref{eqn:creaselagrangian}. The integral over $AdS_2$ diverges but is
regularised to $-2\pi$. Using the equations of motion and the BPS
condition~\eqref{BPS}, the action reduces to
\begin{align}
  S_\text{M2} = \frac{4 N}{\pi} (-2\pi) \int_{\epsilon}^{t_\text{max}}
  \frac{(1+t^2) \diff t}{t^3 \sqrt{1+t^2 - J_\varphi^2 t^4}}\,.
\end{align}
The integral over $t$ gives
\beq
  -\frac{\sqrt{1+t^2-J_\varphi^2 t^4}}{2 t^2} 
  - \frac{1}{4} \arctanh \frac{2 \sqrt{1+t^2-J_\varphi^2 t^4}}{2 + t^2} \,.
\eeq
This vanishes at $t = t_\text{max}$. At $t = \epsilon$ the first term gives an
$\epsilon^{-2}$ divergence corresponding to the usual area divergence; it can be
safely discarded (or removed by an appropriate counterterm). The second term
gives an important $\log\epsilon$ divergence
\bal
  S_\text{M2} &=
  -2N \arctanh\left( 1 - \frac{1}{8} (1 + 4J_\varphi^2) \epsilon^4 +
  \dots \right)
  = N \log\frac{(1+4J_\varphi^2) \epsilon^4}{16}+ \dots
\\&= N \log\frac{\epsilon^4}{16\cos^2(\phi/2)}+ \dots
\label{eqn:BPSactiondiv}
\eal
Note that the coefficient of $\log\epsilon$ matches
the expected conformal anomaly for the sphere $4N + \ho[0]{N} = 4c_1 -
2a_2$~\eqref{anomalycoef}. It must also appear in the form of
$\log{\epsilon/R}$, and reproduces the expected behavior~\eqref{spherevev}.
Subtracting the action for the sphere then gives a finite
quantity, from which we can read the BPS potential
\begin{align}
\label{holo-pot}
  U^{(N)}(\phi,\phi) =
  \frac{N}{\pi} \log{\cos\frac{\phi}{2}}\,.
\end{align}

\subsection{Non-BPS case}

For generic $J_\varphi$ and $J_\vartheta$, the equations of motion~\eqref{eqn:creaseeom2} 
can be brought into canonical elliptic form by performing the change of variables
\begin{align}
  y^2(t)= \frac{(t_\text{max}^2 - t^2)(1+\tau_2)}{(t_\text{max}^2 - \tau_2)(1+t^2)}\,,
  \label{eqn:ellipticchangeofvars}
\end{align}
after which the equations are
\begin{equation}
  \begin{aligned}
    \diff \vartheta &=
    \frac{4 J_\vartheta}{J_\varphi \sqrt{(t_\text{max}^2-\tau_1)(1+\tau_2)}}
    \frac{\diff y}{\sqrt{(1-y^2)(1-y^2/y^2(\tau_1))}}\,,\\
    \diff \varphi &=
    -\frac{1}{\sqrt{(t_\text{max}^2-\tau_1)(1+\tau_2)}}
    \frac{\diff y}{\sqrt{(1-y^2)(1-y^2/y^2(\tau_1))}}
    \left[ 1 - \frac{(1+\tau_2)(1+t_\text{max}^2)}{1+\tau_2 - (\tau_2-t_\text{max}^2) y^2}\right].
    \label{eqn:creaselegendreform}
  \end{aligned}
\end{equation}
These match the definition of elliptic integrals of the first and third kind
\begin{equation}
  \begin{aligned}
    \eF(\psi|k^2) &= \int_0^{\sin{\psi}} \frac{\diff y}{\sqrt{(1-y^2)(1-k^2 y^2)}}\,,\\
    \ePi(n;\psi|k^2) &= \int_0^{\sin{\psi}} \frac{\diff y}{(1-n
      y^2)\sqrt{(1-y^2)(1-k^2 y^2)}}\,.
  \label{eqn:ellipticintegrals}
  \end{aligned}
\end{equation}
The parameters appearing in their definition can be read from above, they are
\beq
\psi = \arcsin\sqrt{\frac{(t_\text{max}^2 - t^2)(1+\tau_2)}{(t_\text{max}^2 - \tau_2)(1+t^2)}}\,, \qquad
k^2 = \frac{(t_\text{max}^2 - \tau_2)(1+\tau_1)}{(t_\text{max}^2 - \tau_1)(1+\tau_2)}\,, \qquad
n = \frac{\tau_2-t_\text{max}^2}{1+\tau_2}\,.
\label{eqn:ellipticparams}
\eeq

Imposing the gluing condition~\eqref{eqn:creasebc2}, we obtain
\beq
\vartheta(t)=
-\frac{4 J_\vartheta}{J_\varphi} \frac{\eF(\psi|k^2)}{\sqrt{(t_\text{max}^2-\tau_1)(1+\tau_2)}}\,,
\qquad
\varphi(t) =
\frac{\pi}{2} +
\frac{\eF(\psi|k^2) - (1+t_\text{max}^2) \ePi(n;\psi|k^2)}{\sqrt{(t_\text{max}^2-\tau_1)(1+\tau_2)}}\,.
  \label{eqn:creasesol}
\eeq
At $t = t_\text{max}$, the elliptic integrals vanish ($\psi = 0$), so the boundary
conditions are manifestly respected.

What is not manifest in these expressions is that $\varphi$ and $\vartheta$
are real functions. The behavior of the roots $\tau_1, \tau_2$ are captured by the
discriminant, they are real and negative if
\begin{align}
  J_\varphi^2 \le -\frac{2}{27} \left[1- 36 J_\vartheta^2 - (1+12J_\vartheta^2)^{3/2} \right]\,,
  \label{eqn:discriminant}
\end{align}
which in particular includes the BPS case $J_\varphi = \pm 2J_\vartheta$, otherwise they are
complex conjugate.
When they are real, taking $\tau_1 < \tau_2$ makes all the quantities real,
including $\varphi$, $\vartheta$. When they are complex conjugate, exchanging
$\tau_1$ and $\tau_2$ is a symmetry so $\varphi$ and $\vartheta$ must be real as well.

\begin{SCfigure}[][tb]
  \centering
  \begin{subfigure}[t]{0.5\textwidth}
  \centering
  \includegraphics[scale=1]{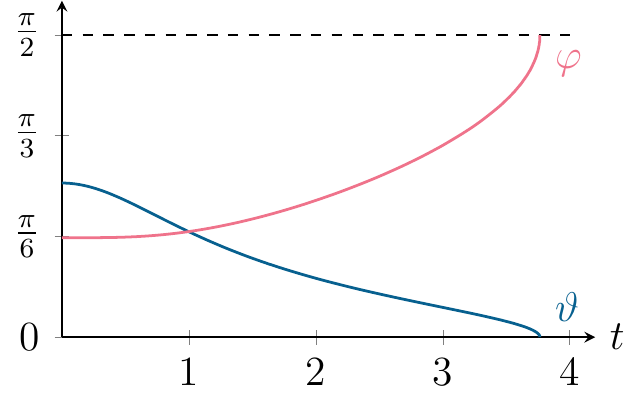}
  \end{subfigure}
  \caption[Example of holographic solution for the crease]
  {Example of solution to the equations of motion~\eqref{eqn:creaseeom2}
  for $\phi = \pi/3$ and $\theta = \pi/2$. $\vartheta(t)$ and $\varphi(t)$ are
  given analytically in~\eqref{eqn:creasesol}, they depend on the parameters
  $J_\varphi = 0.262$, $J_\vartheta = -0.210$ which are found numerically. In this example
  $t_\text{max} = 3.76$.}
  \label{fig:exampleholo}
\end{SCfigure}

Evaluating~\eqref{eqn:creasesol} at $t=0$ relates $\phi$ and $\theta$
to $J_\varphi$ and $J_\vartheta$, but unlike the BPS case, these relations cannot be inverted in terms 
of fundamental functions. Instead we do this perturbatively below around $\phi=\theta$ (near-BPS limit) and 
around $\phi\sim\pi$ and $\theta=0$ (antiparallel limit). Of course, it can also be done 
numerically as in Figure~\ref{fig:exampleholo}.

We now turn to calculating the action. Using the equations of
motion~\eqref{eqn:creaseeom2} and again regularising the volume of $AdS_2$, the
action~\eqref{eqn:creaselagrangian} reduces to
\begin{align}
  S_\text{M2} = \frac{4 N}{\pi} (-2\pi) \int_{\epsilon}^{t_\text{max}}
  \frac{(1+t^2)^{3/2} \diff t}{t^3 \sqrt{(1+t^2)^2 - 4 J_\vartheta^2 t^4 - J_\varphi^2
  t^6}}\,.
  \label{eqn:creaseaction}
\end{align}

Using the change of variables~\eqref{eqn:ellipticchangeofvars}, it too can 
be expressed in terms of elliptic integrals 
\bal
&\int \frac{\diff y}{(1- n' y^2)^2 \sqrt{(1-y^2)(1-k^2 y^2)}}
=
\frac{n'^2 y}{1-n' y^2} \frac{\sqrt{(1-y^2)(1-k^2y^2)}}{2 (k^2-n')(n'-1)}
\\&\hskip1cm
 - \frac{n' \eE(\psi|k^2) + (k^2-n') \eF(\psi|k^2) + \left( n'(n'-2)+k^2(2n'-3) \right)
\ePi(n';\psi|k^2)}{2(k^2-n')(1-n')}\,,
\label{eqn:creaseactionelliptic}
\eal
where $n' = 1/y^2(0)$ differs from $n$ in~\eqref{eqn:ellipticparams} and
$\eE$ is the elliptic integral of the second kind
\begin{align}
  \eE(\psi|k^2) = \int_0^{\sin{\psi}} \diff y \sqrt{\frac{1- k^2 y^2}{1-y^2}}\,.
\end{align}
At $t=0$ both the first term and $\ePi$ diverge. This is the same divergence we
encountered in the BPS case~\eqref{eqn:BPSactiondiv}: to see it, it is easier to
go back to~\eqref{eqn:creaseaction} and expand for small $t$, which gives
\begin{align}
  -8N \int_{\epsilon}^{t_\text{max}} \frac{\diff t}{t^3}
  \left[ 1 + \frac{t^2}{2} + \ho[4]{t} \right]
  = -8N \left [\frac{1}{2 \epsilon^2}
  - \frac{1}{2} \log{\epsilon} \right]
  + \fin\,.
  \label{eqn:creaseactionlogeps}
\end{align}
Notice that both terms are independent of $J_\vartheta, J_\varphi$, so they are independent of the
angles and do not contribute to the potential $U(\phi,\theta)$.

The rest of the integral~\eqref{eqn:creaseactionelliptic} is UV finite and
calculates the potential $U(\phi,\theta)$ (up to the prefactor
$4N/\pi$). 
The dependence on the angles is implicit, since
all the variables and roots of the cubic polynomial~\eqref{eqn:creasepoly}
depend on $J_\varphi, J_\vartheta$, and these are related to $\phi$, $\theta$ by
imposing the boundary conditions at $t=0$ on the solution~\eqref{eqn:creasesol}.

\begin{SCfigure}[][h]
  \centering
  \begin{subfigure}[t]{0.6\textwidth}
  \centering
  \includegraphics{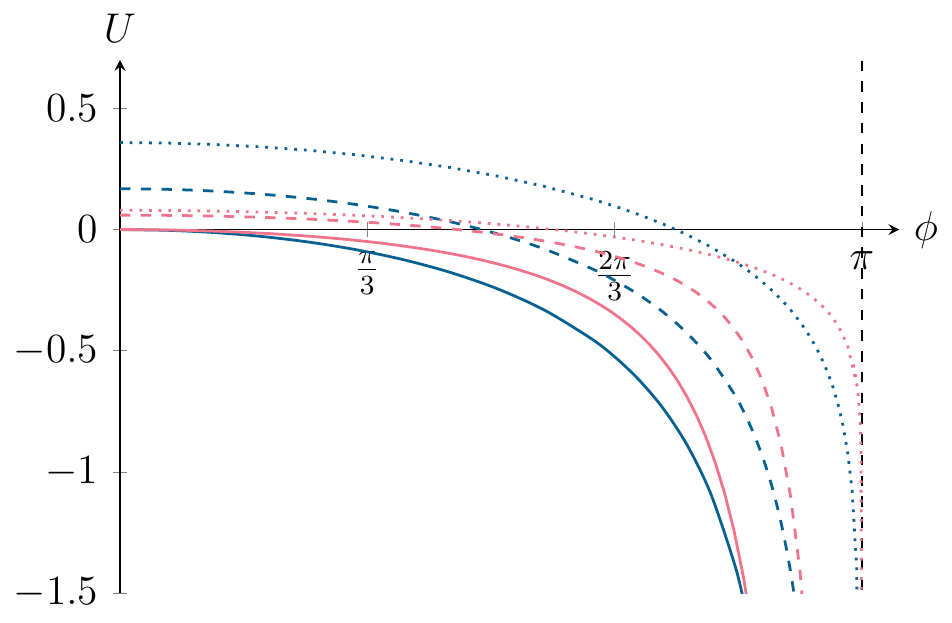}
  \end{subfigure}
  \caption[{Potential $U(\phi,\theta)$ at large $N$}]
  {In blue, the potential $U(\phi,\theta)/N$ at large $N$ obtained numerically
    from~\eqref{eqn:creaseaction} for different values of $\theta$: $\theta = 0$ (solid),
    $2\pi/3$ (dashed) and $\pi$ (dotted). Notice the divergence near
    $\phi = \pi$, which is captured by the antiparallel planes limit \eqref{eqn:antiparallelpotapprox}.
    The potential at $N=1$ for the same values of $\theta$
    \eqref{eqn:potentialabelian} is in red.
  }
  \label{fig:potential}
\end{SCfigure}

It is easy to evaluate the potential for specific $\phi$ and $\theta$ numerically, as
in Figure~\ref{fig:potential}. 

\tikzexternalenable
\tikzexternaldisable

\begin{figure}[htb]
  \centering
  \begin{subfigure}[t]{0.32\textwidth}
    \centering
    \includegraphics{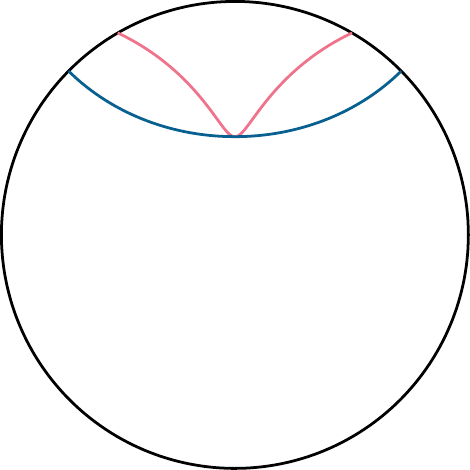}
    \caption{$\theta = \pi/3$}
  \end{subfigure}
  \begin{subfigure}[t]{0.32\textwidth}
    \centering
    \includegraphics{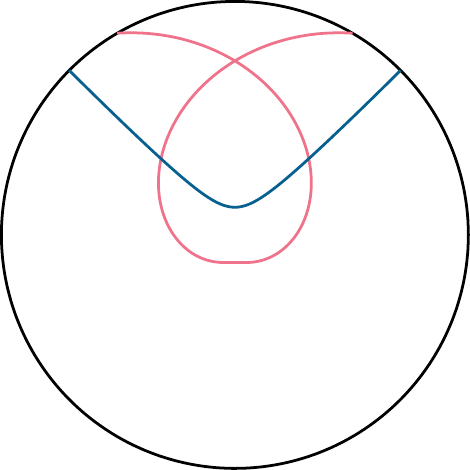}
    \caption{$\theta = 7\pi/3$}
  \end{subfigure}
  \begin{subfigure}[t]{0.32\textwidth}
    \centering
    \includegraphics{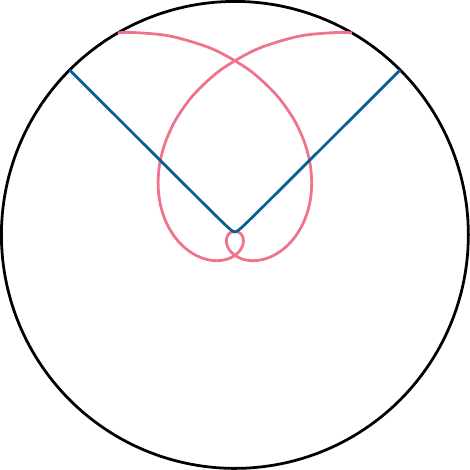}
    \caption{$\theta = 13\pi/3$}
  \end{subfigure}
  \caption{Example of solutions to the equations of motion with nontrivial
    windings for $\phi = \pi/2$ and $\theta=(\pi/3\mod2\pi)$. The radial direction is $(1+t^2)^{-1/2}$, with
  the black circle corresponding to the conformal boundary.  The blue curve is
$\varphi(t)$ and the red curve is $\vartheta(t)$.}
  \label{fig:winding}
\end{figure}

The analysis thus far restricted to solutions with $0 < \vartheta < \theta/2$, but there are 
solutions that are multiply wound around the $S^1_\vartheta$ circle, where for each branch of the solution 
$0 < \vartheta < \pi n+\theta/2$, see for instance Figure~\ref{fig:winding}. Those
solutions can be described analytically, but they also arise when searching
numerically for $J_\varphi$, $J_\vartheta$. 

We should compare their action to that of the unwound solution. They all have the same behaviour and 
divergence at small $t$, which is removed by the same counterterm. 
Looking at the action density, the difference between the solutions with different windings is finite, 
so they can be viewed as instantonic corrections. But it is more appropriate to treat the entire brane including 
the factor of the area of $AdS_2$ in the action~\eqref{eqn:creaselagrangian}. 
One may be tempted to replace the area by $-2\pi$ as in~\eqref{eqn:creaseaction}, 
but this is inconsistent. Firstly, this amounts to different subtractions for different solutions, so does not form a consistent 
regularisation scheme. Furthermore, because of the negative sign, the more the solution is wound, the more 
dominant it would be. If we are to use a consistent regularisation scheme that renders the unwound solution 
finite, the wound solutions have infinite action and are therefore infinitely suppressed. 

\subsection{Near-BPS expansion}
\label{sec:near-BPS}

Here we study the solutions in a systematic expansion beyond the BPS case in Section~\ref{sec:BPS}.

Expanding~\eqref{eqn:creaseeom2} around $J_\varphi = -2J_\vartheta$ gives 
a series of the form
\beq
\vartheta'=\sum_{n=1}^\infty  \frac{p_n(t,J_\varphi)}{(1+t^2)^{n+1} (1+t^2-J_\varphi t^4)^{n+1/2}}(J_\varphi+2J_\vartheta)^n\,,
\eeq
with some polynomials $p_n$ and likewise for $\varphi'$.

The resulting integrals are all trigonometric functions, which is simply the small $k^2$ 
\eqref{eqn:ellipticparams} expansion of the elliptic integrals. Integrating between $t=0$ and
$t_\text{max}$~\eqref{eqn:umaxBPS}, we find
\begin{equation}
  \begin{aligned}
    \phi &=
    2\arctan{2J_\varphi} + \frac{1}{J_\varphi^2} \left[ -\frac{3 + 8 J_\varphi^2}{1 + 4J_\varphi^2} +
    \frac{3}{2 J_\varphi} \arctan{2J_\varphi} \right]
    \left( J_\varphi + 2J_\vartheta \right) + \ho[2]{(J_\varphi+2J_\vartheta)}\,,\\
    \theta &= 2\arctan{2J_\varphi} +
    \frac{1}{J_\varphi^2} \left[ -\frac{6 + 20 J_\varphi^2}{1 + 4J_\varphi^2} + \frac{3}{J_\varphi}
    \arctan{2J_\varphi} \right] \left( J_\varphi + 2J_\vartheta \right) +
    \ho[2]{(J_\varphi+2J_\vartheta)}\,.
  \end{aligned}
\end{equation}
These series can be inverted to give
\begin{equation}
  \begin{aligned}
    J_\varphi &=
    \frac{\tan(\phi/2)}{2} + \left[ \frac{1}{4 \cos^2(\phi/2)} +
    \frac{\tan^3(\phi/2)}{6(\phi-2\tan(\phi/2))} \right]
    \left( \phi - \theta \right)
    + \ho[2]{(\phi-\theta)}\,,\\
    J_\vartheta &=
    -\frac{\tan(\phi/2)}{4} -
    \left[ \frac{1}{8 \cos^2(\phi/2)} +
    \frac{\tan^3(\phi/2)}{6(\phi-2\tan(\phi/2))} \right]
    \left( \phi - \theta \right)
    + \ho[2]{(\phi-\theta)}\,.
  \end{aligned}
\end{equation}

Expanding the action in series also gives trigonometric functions. The lowest
order is the BPS case~\eqref{eqn:BPSactiondiv}, and 
higher-order terms in the $J_\varphi-2J_\vartheta$ expansion calculate the
near-BPS corrections to the potential. As we already proved, no further 
$\log\epsilon$ terms arise and we only find new power-like divergences, which 
can be discarded. The potential is then
\bal
U^{(N)}(\phi,\theta) = 
\frac{N}{\pi} \bigg\{&
\log{\cos\frac{\phi}{2}}
-\frac{\tan(\phi/2)}{2} ( \phi - \theta)
-  \left[ \frac{1}{\cos^2(\phi/2)}
+\frac{4\tan^3(\phi/2)}{3(\phi-2\tan(\phi/2))} \right]\frac{( \phi - \theta)^2}{8}
\\
&{} - 
\frac{\tan(\phi/2)}{\cos^2(\phi/2)}\left[ \frac{1}{3} +
\frac{2\tan(\phi/2)}{\phi-2\tan(\phi/2)} 
+\frac{10 (13+\cos{\phi}) \tan^2(\phi/2)}{27(\phi-2\tan(\phi/2))^2}
\right.\\
&\hskip1in + \left. 
\frac{280\sin^2(\phi/2) \tan^3(\phi/2)}{81(\phi-2\tan(\phi/2))^3}
\right] (\phi - \theta)^3
+ \cdots \Big\}\,.
  \label{eqn:genpotentialholo}
\eal

\subsection{Antiparallel planes limit}

The limit $\pi - \phi=\delta \ll 1$ is not much simpler than the general case but is
interesting because it calculates the potential between antiparallel planes. It
corresponds to taking $J_\varphi \to \infty$. To retain nonzero $\theta$ we
should take a double scaling limit $J_\vartheta \to \infty$ keeping $q =
J_\vartheta/J_\varphi^{2/3}$ fixed.  The polynomial \eqref{eqn:creasepoly} then
becomes (with $t=s/J_\varphi^{1/3}$)
\begin{align}
  1- 4 q^2 s^4 - s^6 = -(s^2 - \tau_1)(s^2 - \tau_2)(s^2 - s_\text{max}^2)\,.
\end{align}
To leading order in $J_\varphi$, we have
\begin{align}
  \vartheta =
  - 4 q \int_0^{s_\text{max}} \frac{s\, \diff s}{\sqrt{1- 4 q^2 s^4 - s^6}}
  + \ho[-2/3]{J_\varphi}\,.
  \qquad
\end{align}
Expressing the integral in terms of elliptic integrals~\eqref{eqn:ellipticintegrals} gives
\begin{align}
  \theta = \frac{8 q}{\sqrt{s_\text{max}^2 - \tau_1}}
  \eF\left(\arcsin\frac{s_\text{max}}{\sqrt{s_\text{max}^2 - \tau_2}} \,\bigg|\,
  \frac{s_\text{max}^2 - \tau_2}{s_\text{max}^2 - \tau_1}\right)
  + \ho[-2/3]{J_\varphi} \,.
  \label{eqn:antiparalleltheta}
\end{align}
This is an implicit relation between $q$ and $\theta$.
Similarly for $\phi$, expanding~\eqref{eqn:creaseeom2} to leading order in $J_\varphi$
gives 
(where all the elliptic integrals have the same arguments as in \eqref{eqn:antiparalleltheta})
\begin{align}
  \delta=\pi-\phi =  2J_\varphi^{-1/3} \left[ 
    \frac{\tau_1}{\sqrt{s_\text{max}^2 - \tau_1}} \eF + \sqrt{s_\text{max}^2 - \tau_1} \eE
  \right] + \ho[-1]{J_\varphi}\,.
  \label{eqn:antiparallelphi}
\end{align}
The action at leading order is
\begin{align}
  S_\text{M2} = 4 N J_\varphi^{2/3} 
  \left(
    \frac{\tau_1}{\sqrt{s_\text{max}^2 - \tau_1}} \eF + \sqrt{s_\text{max}^2 - \tau_1} \eE
  \right)
  + \ho[0]{J_\varphi}\,.
\end{align}
Unlike~\eqref{eqn:creaseactionelliptic}, here there are no divergences to treat
because the $\log{\epsilon}$ divergence is subleading in $J_\varphi$.
Using~\eqref{eqn:antiparallelphi} to solve for $J_\varphi$, we find
\begin{align}
  U^{(N)}(\pi-\delta,\theta) = -\frac{8 N}{\pi \delta^2} \left( 
    \frac{\tau_1}{\sqrt{s_\text{max}^2 - \tau_1}} \eF + \sqrt{s_\text{max}^2 - \tau_1} \eE
  \right)^3
\,.
  \label{eqn:antiparallelpot}
\end{align}
The dependence on $\theta$ is implicit. For small $\theta$, $q$ is small
and we can invert~\eqref{eqn:antiparalleltheta} explicitly
\begin{align}
  \theta = \frac{4 \sqrt{\pi} \Gamma( 1/3)}{3 \Gamma(5/6)} q
  -\frac{16q^3}{3} + \ho[5]{q}\,.
\end{align}
Expanding~\eqref{eqn:antiparallelpot} at small $q$ and inverting $q(\theta)$
gives
\begin{equation}
  \begin{aligned}
    U^{(N)}(\pi-\delta,\theta) =
    -\frac{8 N \sqrt{\pi} \Gamma({2}/{3})^3}{\Gamma(1/6)^3 \delta^2}
    &\left[ 1 - \frac{3\sqrt{3}}{8\pi} \theta^2
      + \left( 1 - \frac{3^{5/2}  \Gamma(2/3) \Gamma(5/6)^4}{2^{7/3} \pi^3}
      \right) \frac{9 \theta^4}{64 \pi^2} \right.\\
      &\  + \dots \Big]
      + \frac{2N}{\pi} \log\delta + \dots\,.
  \end{aligned}
  \label{eqn:antiparallelpotapprox}
\end{equation}
For $\theta=0$, this matches the original calculation
of the antiparallel planes~\eqref{eqn:potentialantiparallelplanes} 
\cite{maldacena:1998im}. Here we have a perturbative expansion that can easily be continued to 
higher orders.

The free field result \eqref{eqn:potentialabelian} is exact in $\phi$ and $\theta$, but 
expanding it around $\phi=\pi$ yields
\beq
U^{(1)}(\pi-\delta,\theta) =
\frac{1}{2\pi} \log\sin\delta/2 - \frac{\cos\delta + \cos\theta}{8\pi
\sin^2\delta/2}
=-\frac{1+\cos\theta}{2\pi\delta^2}+\frac{1}{2\pi}\log\delta+\dots
\eeq
Like \eqref{eqn:antiparallelpotapprox}, this too has a double pole at $\delta=0$ and a logarithmic correction.

Note the coefficient of the 
$\log\delta$ term in the holographic expression~\eqref{eqn:antiparallelpotapprox} 
is double that of the $\log\cos\phi/2$ part for the BPS expression 
in~\eqref{eqn:genpotentialholo}. That implies that for $\theta=\pi-\delta$, the series above 
sums up to $\sim\delta^2\log\delta$, to cancel the power law divergence and fix the 
prefactor of the log.

\section{Lightlike crease}
\label{sec:lightlike}

A famous limit of the cusped Wilson loop is when the two rays approach the lightcone in Minkowski 
space. The resulting anomaly is proportional to the boost parameter between the lines, with a coefficient 
known as the universal cusp anomalous dimension. It governs hard processes in QCD 
\cite{Korchemsky:1988si, Korchemsky:1992xv} and also plays a crucial role in $\cN=4$ SYM, 
where it is captured by the spinning string solution \cite{Gubser:2002tv}, the cusped Wilson 
loop in Minkowski space \cite{Kruczenski:2002fb}, and served as the cornerstone for the analysis 
of scattering amplitudes in that theory including the BDS ansatz \cite{Bern:2005iz} and the Wilson loop 
scattering amplitude duality \cite{Alday:2007he}.

We present here analog calculations for a lightlike crease, starting by the analytic continuation of 
the solution in Section~\ref{sec:AdS} to imaginary angle $\phi$ 
and then writing down solutions in lorentzian $AdS$ including 
the analogue of the 4-cusp solution, which turns out to be two lightcones joined along a circular lightlike 
crease.

A different adaptation of the 4-cusp solution to surface operators was studied in~\cite{Bhattacharya:2016ydi}. 
We comment on the differences below.

\subsection{Analytic continuation}

We first look at the solution in Section~\ref{sec:AdS} and find a limit where $\phi$ diverges. 
The equation for $\varphi$~\eqref{eqn:creaseeom2} can be integrated over the zeros of the 
polynomial~\eqref{eqn:poly} in the denominator, and its integral converges
unless the zeros are degenerate. The latter is determined by the vanishing of the
discriminant~\eqref{eqn:discriminant}
\begin{align}
J_\varphi^2 = -\frac{2}{27} \left[1- 36 J_\vartheta^2 \pm(1+12J_\vartheta^2)^{3/2} \right]\,.
\end{align}
We want to get imaginary $\phi$, so should take the negative sign and as the 
result should not depend on $\theta$, we can take $J_\vartheta=0$ and
\begin{align}
  J_\varphi^2 = -\frac{4}{27}\,.
\end{align}
Now we get that $\varphi$ is given by~\eqref{eqn:creaseeom2}
\begin{align}
\varphi' = \pm\frac{2 it^3}{(t^2 + 3) \sqrt{(1+t^2)(3+4t^2)}}\,,
\end{align}
and its integral up to the root $t^2=-3$ gives $(\pi-\phi)/2$. With a regulator $\delta$, we have
\begin{align}
\label{imaginaryphi}
\phi =
\pi+4 i \int_0^{\sqrt{3} i -\delta i}\frac{t^3 \diff t}{(t^2 + 3) \sqrt{(1+t^2)(3+4t^2)}}\,.
\end{align}
The real part of the integral cancels the factor $\pi$, while the imaginary part
diverges logarithmically as $\delta \to 0$. 
Its coefficient is the residue at $t =\sqrt{3} i$, so we get
\begin{align}
\phi =-{i}{\sqrt{2}} \log |\delta| + \dots \to i \infty\,.
\end{align}

To evaluate the action in that limit, we take the analytic continuation
of~\eqref{eqn:creaseaction}. The integral becomes
\begin{align}
  S_\text{M2} =
  \frac{4N}{\pi} (-2\pi) \int_\epsilon^{i \sqrt{3}-i \delta}
  \frac{3^{3/2}(1+t^2)^{3/2} \diff t}{t^3 (3+t^2) \sqrt{3+4t^2}}\,.
\end{align}
Again the pole gives a logarithmic divergence. Taking the residue and comparing
against $\phi$ we find
\begin{align}
\label{delta-action}
  S_\text{M2} =
  8 N \frac{\sqrt{2}}{3^{3/2}} \log|\delta| + \dots
  \to -\frac{8N |\phi|}{3^{3/2}}\,, \qquad
  \phi \to i \infty\,.
\end{align}
As in the case of the lightlike cusp~\cite{Kruczenski:2002fb}, we find a result proportional to the boost angle $|\phi|$ 
and the coefficient may be seen as the analogue of the universal cusp anomalous dimension.

The linear growth of the expectation value at large boost is also present in the
free theory. There the analytic continuation is trivial, simply
take~\eqref{eqn:potentialabelian} in the limit $\phi \to i \infty$ to obtain
\begin{align}
  U^{(1)} \to \frac{|\phi|}{2\pi} + \dots
\end{align}
We do not know whether there is really a meaning to this expression as a potential and to what extent we should 
be analytically continuing the solution for the compact crease rather than the infinite one. To address these points 
we turn now to study the system directly in lorentzian signature.

\subsection{Solutions in lorentzian $AdS$}
\label{sec:lorentz}

We now look for solutions to represent the infinite crease in Minkowski space. We use the 
metric of lorentzian $AdS_4$ subspace of $AdS_7\times S^4$
\beq
\diff s^2=\frac{4L^2}{z^2}(\diff z^2+(\diff x^1)^2+(\diff x^2)^2-\diff t^2)
=\frac{4L^2}{z^2}(\diff z^2+(\diff x^1)^2+\diff\xi^2-\xi^2\,\diff\chi^2)\,.
\eeq
As world-volume coordinates we take (for $|x^2|>|t|$) 
$\xi=\sqrt{(x^2)^2-t^2}$, $\chi=\arccoth(x^2/t)$ and $v$ and the ansatz
\beq
z=u(\chi)\xi\,,\qquad
x^1=\sqrt{1+u(\chi)^2}\,v\,.
\eeq
The scaling of $z$ with $\xi$ is natural because of dilatation invariance of the crease. The scaling of 
$x^1$ with $\sqrt{1+u^2}$ may seem unnatural at first, but it is required to have a full 1d conformal 
symmetry $SO(2,1)$ as is discussed in Appendix~\ref{app:poincare}.
This seems to be one difference from the solution in \cite{Bhattacharya:2016ydi} and there should 
be more solutions interpolating between the two.

Given this, the lagrangian is
\beq
\cL=8T_\text{M2}L^3
\frac{\sqrt{(1+u^2)(1+u^2-u'^2)}}{\xi^2u^3}\,.
\label{lorentzlag}
\eeq
Ignoring $8L^3/\xi^2$, the conserved energy is
\beq
E=\frac{(1+u^2)^{3/2}}{u^3\sqrt{1+u^2-u'^2}}\,.
\eeq
Inverting this gives
\beq
\chi=\int \frac{E u^3\,\diff u}{\sqrt{(1+u^2)(E^2u^6-(1+u^2)^2)}}\,.
\eeq
In these coordinates we can find the solutions that are the analytic continuation of those in Appendix~\ref{app:poincare} 
for finite boost angle. The solutions we get do not cover the entire world volume and they require continuation to 
$|x^2|\leq|t|$, where we take instead the metric
\beq
\label{minkmet}
\diff s^2
=\frac{4L^2}{z^2}(\diff z^2+(\diff x^1)^2-\diff\xi^2+\xi^2\,\diff\chi^2)\,,
\eeq
where now $\xi=\sqrt{t^2-(x^2)^2}$ and $\chi=\arccoth(t/x^2)$.

To study solutions in this patch \eqref{minkmet}, the ansatz is
\beq
z=u(\chi)\xi\,,\qquad
x^1=\sqrt{u(\chi)^2-1}\,v\,,
\eeq
and the lagrangian becomes
\beq
\label{lorentzlag2}
\cL=8T_\text{M2}L^3
\frac{\sqrt{(u^2-1)(u^2-1-u'^2)}}{\xi^2u^3}\,.
\eeq

The simplest solution to the equations of motion has constant $u$, where imposing that in the 
Euler-Lagrange equation for \eqref{lorentzlag2} gives the relation $u^2=3$. In the original 
coordinate patch this is $u=\pm i\sqrt3$, which also arises near the singularity of 
\eqref{imaginaryphi}, where the solution becomes stationary.

For this solution in the new patch, the action becomes
\beq
S_\text{M2}=\frac{4N}{3\sqrt3\,\pi}\int
\frac{\diff v\,\diff\xi\,\diff\chi}{\xi^2}\,.
\label{eqn:lightlikecrease}
\eeq
This is proportional to the area of $AdS_2$ times the extent of the coordinate $\chi$.
Comparing to \eqref{delta-action}, we see that they agree,  if we replace the $v$, $\xi$ integral with 
the integral over global euclidean $AdS_2$ with regularised area $-2\pi$.

\subsection{Global solutions}
\label{sec:AM}

The lightlike crease solution above is obtained in the
Poincar\'e patch and can be embedded in global $AdS$. For the lightlike cusp
solution of~\cite{Kruczenski:2002fb}, the analysis of the global structure of
the solution revealed that the lightlike cusp is in fact related by a conformal
transformation to the lightlike Wilson loop with 4-cusps~\cite{Alday:2007hr}.
Here we find that the single lightlike crease is conformally related to a surface
consisting of two lightcones glued along a crease.

Following \cite{Kruczenski:2002fb,Alday:2007hr}, we can embed our solution into
global $AdS$ by writing it in terms of the embedding coordinates
\bal
X_{-1}&=\frac{1}{2z}(L^2+z^2-t^2+(x^i)^2)\,,\quad
&X_{0}&=\frac{Lt}{z}\,,\\
X_{6}&=-\frac{1}{2z}(-L^2+z^2-t^2+(x^i)^2)\,,\quad
&X_{i}&=\frac{Lx^i}{z}\,,\quad
\eal
as
\beq
X_0^2-X_2^2=\frac{L^2}{3}\,,
\qquad
X_3=X_4=X_5=0\,,
\eeq
and then $X_{-1}^2-X_1^2-X_6^2=2L^2/3$. This can be written homogenously  as
\beq
X_{-1}^2-X_1^2-X_6^2=2X_0^2-2X_2^2\,,
\qquad
X_3=X_4=X_5=0\,.
\eeq

Conformal transformations are given by rotations of the embedding coordinates,
and we can now consider the rotated solution
\beq
\label{hom}
X_0^2-X_1^2-X_2^2=2X_{-1}^2-2X_6^2\,,
\qquad
X_3=X_4=X_5=0\,.
\eeq
In the new Poincar\'e patch, this new solution is
\beq
z
=\sqrt{\frac{3}{2}(t^2-r^2)}\,,
\eeq
with $r=\sqrt{(x^1)^2+(x^2)^2}$. For $z=0$, this is a two-dimensional lightcone $t=r$.

We can write its action as
\begin{align}
S_\text{M2}
= \frac{4 N}{3\pi} \sqrt{\frac{2}{3}}
\int \frac{\sinh\rho}{z}\diff z\, \diff \rho\, \diff v\,,
\end{align}
where $\rho=\arccoth(t/r)$ and $v$ the angle in the $(x^1,x^2)$ plane parametrising an $AdS_2$ at fixed $z$. 
The integral over $AdS_2$ is regularised to $-2\pi$ and
the integral over $z$ diverges as $\log{z}$.
This agrees with the analytic continuation~\eqref{delta-action} if we identify
the $\log{z}$ divergence with $|\phi|/\sqrt{2}$.

To obtain a compact surface, we start
again with \eqref{hom}, swap $X_6$ with $X_3$ and map 
$X_{0}\to(X_0+X_{-1})/\sqrt2$ and 
$X_{-1}\to(X_{-1}-X_{0})/\sqrt2$. We find
\beq
2X_{-1}^2-X_{0}^2\to 
\frac{X_{-1}^2+X_0^2-6X_{-1}X_0}{2}=2X_3^2-X_1^2-X_2^2\,,
\qquad
X_4=X_5=X_6=0\,.
\eeq
This can be recast as
\beq
\frac{z^2}{3} = \frac{1}{2} (t-L)^2 - (x^3)^2\,, \qquad
r^2 = 4L^2 + 2(x^3)^2 - \frac{1}{2} (t-3L)^2\,.
\label{eqn:lightlike}
\eeq
Setting $z=0$ gives the surface on the boundary parametrised by $t$ (and angle $v$) with two branches
\begin{align}
  x^3 = \pm \frac{L-t}{\sqrt{2}}\,, \qquad
  r = \frac{L+t}{\sqrt{2}}\,.
\end{align}
The physical domain is $-L < t < L$, so $r, |x^3| < \sqrt{2} L$.
Each branch parametrises a lightcone emanating from $t=-L$ and $x^3 = \pm\sqrt{2} L$ 
and merging along a crease at $t=L$, $x^3=0$ and $r =
\sqrt{2}L$.

It's convenient to parametrise the solution by $z$, $x_3$ and $v$, such that 
the action is
\begin{align}
  S_\text{M2} = 
  \frac{4 L N}{\sqrt{3} \pi}
  \int \frac{\diff z\, \diff x_3\, \diff v}
  {z^2 \sqrt{z^2 + 3 x_3^2}}\,.
\end{align}
There are two sources of divergence. First when $z$ goes to $0$, which is the
usual divergence from the volume of $AdS$, and then in addition when also $x_3
\to 0$, which is associated with the crease singularity.
Introducing a cutoff $z > \epsilon$ regularises both.
To perform the integral, notice that the M2-brane closes in $AdS$ when
$r = 0$, which sets the range of $x_3$ and $z$ to be
\begin{align}
  x_3^2 < \frac{(z-2\sqrt{3}L)^2 - 6L^2}{3}\,, \qquad
  z < \sqrt{6}(\sqrt{2}-1)L\,.
\end{align}
Integrating we find
\begin{align}
  S_\text{M2} = -\frac{16 N}{3} \frac{L}{\epsilon} \log{\epsilon} + \fin.
\end{align}
This result is peculiar and is in contradiction to our assertion in Section~\ref{sec:creaseanomaly} 
that creases do not suffer from $\epsilon^{-1} \log{\epsilon}$ divergences. It deserves further 
exploration.

An alternative regularisation which preserves the $AdS_2$ symmetry is to write
the lorentzian $AdS_7$ as a foliation over $AdS_2 \times dS_4$, as
in the case of the compact crease in Section~\ref{sec:setup}. 
This is the analytic continuation in $\nu$ of~\eqref{ads-metric1}. 
Keeping only the timelike coordinate from the de Sitter component, the metric is
\beq
ds^2=L^2\left[d\nu^2+\cos^2\nu(\diff \rho^2 + \sinh^2 \rho\, \diff v^2)
- \sin^2 \nu\,\diff \varphi^2\right].
\eeq
The boundary of space is at $\varphi\to\pm\infty$ or $\rho\to\infty$. 
Taking $\nu(\alpha)$ the action is
\beq
\cL=8T_\text{M2}L^3
\cosh\rho\cos^2\nu\sqrt{\sin^2\nu-\nu'^2}\,.
\eeq
Looking for constant $\nu$ solutions gives $\sin^2\nu=1/3$ and regularising the $AdS_2$ area leads to 
the action
\begin{align}
  S_\text{M2} = -\frac{8 N}{3\sqrt3} \int d\varphi,
\end{align}
which again agrees with the analytic continuation of the
potential~\eqref{delta-action} and the action of the lightlike
crease~\eqref{eqn:lightlikecrease}.

\section{Defect CFT}
\label{sec:dCFT}

As the crease preserves part of the conformal group, it is natural to study it from a defect CFT 
perspective \cite{Cardy:1991tv, McAvity:1995zd, Liendo:2012hy, Gaiotto:2013nva,
Billo:2016cpy, Antunes:2021qpy}. 
In particular, we use these techniques to find 
expressions for the potential as an expansion around BPS configurations, 
including quadratic order around the plane, producing~\eqref{eqn:potentialnearplane} 
and linear order around the $\theta=\phi$ cusp, resulting in~\eqref{eqn:potentialnearBPS2}.
This approach does not rely on a particular realisation of the $\cN = (2,0)$ theory and 
the surface operators, so is applicable to theories with any $ADE$ algebra and surface operators 
in any representation of the algebra.

The principle at action is very similar to that of the bremsstrahlung function $B(\lambda)$ 
and ``generalised bremsstrahlung function'' $B(\phi,\lambda)$ for cusped Wilson 
loops in $\cN=4$ SYM (with multiple generalisations) 
\cite{correa:2012at, Gromov:2012eu, Fiol:2015spa, Bianchi:2018scb}.

A superconformally invariant surface has a distinguished set of operators restricted to the defect, known as 
the displacement operator multiplet, with conformal primaries:
$\bD^m$ ($m = 3,\dots,6$) for broken translations, $\bO^i$ ($i = 2,\dots,5$) for broken R-symmetries
and $\bQ$ for broken supercharges 
\cite{Drukker:2017xrb, Drukker:2017dgn,  bianchi:2018zpb, Bianchi:2019sxz, Drukker:2020atp}. 
For the spherical surface operator, 
they can be expressed as contact terms for the conservation equations for the energy momentum 
tensor $T^{\mu\nu}$ and R-current $j^{\mu IJ}$ as
\begin{align}
  \partial_\mu T^{\mu \nu}(x_\parallel, x_\perp) V_{S^2} &=
  e^{\nu}_n V_{S^2}[\bD^n(x_\parallel)] \delta^{(4)}(x_\perp)\,,\\
  \partial_\mu j^{\mu i 1}(x_\parallel, x_\perp) V_{S^2} &=
  V_{S^2}[\bO^i(x_\parallel)] \delta^{(4)}(x_\perp)\,.
  \label{eqn:dispop}
\end{align}
There is also a similar equation for broken supersymmetries $\bQ$. 
Here $x_\parallel$ are the coordinates along the surface and 
$x_\perp$ are normal. Furthermore, we use the notation
$V[\bD]$ to denote the insertion of the operator $\bD$ on the defect.
$e^{\nu}_n$ are the vielbein restricted to the normal space at the point
$x_\parallel$.

These operators are special because, as we review below, 
their correlators capture the expectation value of small deformations of the conformal defect. 
We start with the case of deformations of the sphere to
find~\eqref{eqn:potentialnearplane},
and then proceed to study the deformations away from the BPS crease at finite 
$\phi=\theta$ to derive~\eqref{eqn:potentialnearBPS2}.

\subsection{Near sphere expansion}
\label{sec:smallangle}

Consider a surface which is a small deformation of a sphere. This can be
obtained by a local deformation along a normal vector
$\xi_\nu(x_\parallel)$. It can be expressed as a six dimensional integral over the derivative of the 
current $\partial_\mu T^{\mu\nu}$, and including also a local R-symmetry deformation $\omega^i$
leads to a formal expression defining the operator $V_{\xi,\omega}$
\begin{align}
  V_{\xi,\omega} =
  \exp\left[ \int \diff^6 x\left(
  \xi_\nu \partial_\mu T^{\mu\nu} + \omega^i \partial_\mu j^{\mu 1i} \right)\right]
  V_{S^2}\,.
  \label{eqn:deformationV}
\end{align}
Clearly if $\xi$ and $\omega$ are constants, the exponential simply expresses a
global translation/R-symmetry rotation. As we allow $\xi$, $\omega$ to vary
along the sphere, it describes a local deformations that can affect the expectation value. 
We adopt spherical coordinates for
$\bR^6$
\begin{align}
  \diff s^2 =
  \diff r^2 + r^2 \left( \diff u^2 + \cos^2{u}\, \diff v^2
  + \sin^2{u} \left( \diff \theta_3^2 + \cos^2{\theta_3} \left( \diff \theta_2^2
  + \cos^2{\theta_2}\, \diff \theta_1^2 \right) \right)\right),
  \label{eqn:metricr6sph}
\end{align}
and the sphere is located at $\theta_{1,2,3} = 0$, $r=R$. $\xi$, $\omega$ are now 
functions of $u$, $v$, and since $\xi$ is a normal vector, it can have components 
$\partial_{\theta_1}, \partial_{\theta_2}, \partial_{\theta_3}$ and $\partial_{r}$.

We can calculate the expectation value of $V_{\xi,\omega}$ for small $\xi$, $\omega$ 
by expanding the exponential in~\eqref{eqn:deformationV}. 
At the linear order, using~\eqref{eqn:dispop}, we get insertions of 
displacement operators into the surface operator
\begin{align}
  R^2 \int \cos{u} \,\diff u \,\diff v\, 
  \left(\xi_\nu(u,v) e^\nu_nV_{S^2}[ \bD^n(u,v)] + \omega^i(u,v)V_{S^2}[\bO^i(u,v)]\right).
\end{align}
The factor $R^2 \cos{u}$ comes from the metric on the sphere of
radius $R$.

These terms do not contribute to the expectation value since 1-point functions
vanish by symmetry. The first nontrivial contribution is instead at quadratic order, 
where we find pairs of displacement operators from the currents acting on $V$. In addition 
there are contact terms from the currents acting on defect operators. This can be seen as 
delta function contributions to the $\bD\bD$ and $\bO\bO$ OPEs, giving
\begin{align}
  \label{eqn:jVO}
  \nonumber \partial_\mu j^{\mu i 1}(u,v,x^\perp) V_{S^2}[\bO^j(u',v')]
  &= V_{S^2}[\bO^i(u,v) \bO^j(u',v')] \delta^{(4)}(x^\perp)\\
  &\quad + \frac{\alpha}{R^2} V_{S^2}[\mathds{1}] \delta^{ij} \delta^{(4)}(x^\perp)
  \delta(\sin(u-u')) \delta(v-v')\,,\\
  \label{eqn:TVD}
  \nonumber \partial_\mu T^{\mu \nu}(u,v,x^\perp) V_{S^2}[\bD^n(u',v')]
  &= V_{S^2}[e^\nu_m \bD^m(u,v) \bD^n(u',v')] \delta^{(4)}(x^\perp)\\
  &\quad + \frac{\beta}{R^4} V_{S^2}[\mathds{1}] e^\nu_m \delta^{mn} \delta^{(4)}(x^\perp)
  \delta(\sin(u-u')) \delta(v-v')\,.
\end{align}
The contact terms come with unknown prefactors $\alpha$ and $\beta$ and only the 
identity part of the singular OPE is retained, since all other 1-point functions vanish. 
These terms come with appropriate powers fo $R$, on dimensional grounds, so they 
don't appear in the dCFT description of the plane in~\cite{Drukker:2020atp}.

Using~\eqref{eqn:jVO},~\eqref{eqn:TVD} we obtain the change in
expectation value for arbitrary $\xi(u,v)$, $\omega(u,v)$
\begin{equation}
\begin{aligned}
  \vev{V_{\xi,\omega}} - \vev{V_{S^2}} =&\ 
  \frac{R^4}{2} \int \cos u\, \diff u\, \diff v \cos u'\,\diff u'\, \diff v'\,
  \omega^i \omega^j \vev{V_{S^2}[\bO^i(u,v) \bO^j(u',v')]}\\
  & + \frac{R^4}{2} \int \cos u \,\diff u \,\diff v \cos u' \,\diff u' \,\diff v'\,
  \xi^\mu e_\mu^m \xi^\nu e_\nu^n \vev{V_{S^2}[\bD^m(u,v) \bD^n(u',v')]} \\
  & + \frac{1}{2} \vev{V_{S^2}} \int \left( \alpha \omega^2 + \frac{\beta \xi^2}{R^2}\right) \cos u\, \diff u\, \diff v
  + \dots
\end{aligned}
  \label{eqn:Vomegavev}
\end{equation}

The 2-point functions of $\bD$ and $\bO$ that enter this expression are fixed by
their dimensions 3 and 2 respectively and for the spherical surface are
\begin{align}
\label{eqn:2pt}
  \frac{\vev{V_{S^2}[\bD^m(u,v) \bD^n(u',v')]}}{\vev{V_{S^2}}} &=
  \frac{C_\bD}{8\pi^2R^6} \frac{\delta^{mn}}{(1-\sin u \sin u' -
  \cos u\cos u'\cos(v-v'))^3}\,,\\
  \frac{\vev{V_{S^2}[\bO^i(u,v) \bO^j(u',v')]}}{\vev{V_{S^2}}} &=
  \frac{C_\bO}{4\pi^2R^4} \frac{\delta^{ij}}{(1-\sin u \sin u' -
  \cos u\cos u'\cos(v-v'))^2}\,.
\end{align}
The normalisation of $\bD$ is fixed by~\eqref{eqn:dispop}, so
$C_\bD$ and $C_\bO$ on the right hand side are not arbitrary and are part of the dCFT data. 
In fact, for any surface operator in a CFT, it was shown in~\cite{bianchi:2015liz} that $C_\bD$ 
is related to the coefficient $a_2$ entering the
conformal anomaly~\eqref{anomalydens} as $C_\bD=-16a_2$. 
For the $\cN = (2,0)$ theory, supersymmetry implies 
that $a_2 = -c$ and the normalisation of the 
2-point function of $\bO$ is given by $C_\bO=c$~\cite{Drukker:2020atp}.

It is now a simple matter to fix $\alpha$ and $\beta$.
Taking $\xi$ to be any conformal Killing vector, for example 
$\partial_{\theta_1}$, leads to a global rotation which should not change the
expectation value. This fixes the constant $\beta$. We can also fix $\alpha$ by taking
$\omega^{i} = \delta^{i2}$ to generate a constant R-symmetry rotation.
Performing the integrals~\eqref{eqn:Vomegavev} with these values of $\xi$, $\omega$, we find
\begin{align}
  \alpha = \frac{c}{4\pi}\,, \qquad
  \beta = -\frac{3 c}{2 \pi}\,.
\end{align}

With $\alpha$, $\beta$ in hand, we now evaluate~\eqref{eqn:Vomegavev} for the case of
the crease of angle $\phi$, $\theta$ as defined in~\eqref{eqn:screaseparam}
and~\eqref{eqn:sscalarparam}. We take
\begin{align}
  \omega^{i} =
 \begin{cases}
    \theta \delta^{i2}\,, & u>0\\
    0\,, & u<0\\
  \end{cases}\,, \qquad
  \xi^{\mu} e_\mu^m =
  \begin{cases}
    \phi e_{\theta_1}^m = \phi R \sin{u} \delta^{6m}\,, & u>0\\
    0\,, & u<0\\
  \end{cases}\,,
\end{align}
which rotates one of the hemisphere by an angle $\phi$ and the scalar coupling
by $\theta$. Plugging these in~\eqref{eqn:Vomegavev} and performing the integrals we obtain
\begin{align}
  \frac{\vev{V_{\phi,\theta}}}{\vev{V_{S^2}}} =
  1 + \frac{c}{4} (\theta^2-2\phi^2) + \dots
  \label{eqn:Vtheta2}
\end{align}
from which~\eqref{eqn:potentialnearplane} follows directly. This is also in agreement with 
the expansion of the explicit results~\eqref{eqn:potentialabelian} and~\eqref{holo-pot}.

\subsection{Near-BPS expansion}
\label{sec:nearBPS}

The previous subsection relied on the defect CFT description of the deformation of the sphere. 
To study the crease with finite angles $\phi$, $\theta$, we now turn to the
formulation of the defect CFT description of deformations of the crease itself.

It is easiest to derive the formalism for the infinite crease and then use a conformal transformation 
to apply it to the spherical crease. The symmetries of the infinite crease~\eqref{infinitecrease}
are $\sof(2,1) \oplus \sof(3) \oplus\sof(3)$, corresponding respectively to the 1d conformal 
group acting on both half-planes, transverse rotations in $x^{4,5,6}$ and transverse R-symmetry
rotations of $n^{3,4,5}$. Note that from the point of view of symmetries, this
setup is identical to inserting a line operator inside the plane. According to the classification of 
\cite{Agmon:2020pde}, this is a supersymmetric line defect breaking the transverse rotational symmetry.

On the sphere or plane, $\bD^m$ and $\bO^i$ are quartets of a pair of $\sof(4)$ symmetries, broken to 
$\sof(3)$ by the crease. This singles out one displacement $\bD^3$ (for the crease in the $(2,3)$ plane extended 
in the $x^1$ direction) and one R-rotation $\bO^2$ that 
are singlets and can therefore have expectation values. Another way to say this is that these operators are 
no longer primaries and they mix with the identity.
Explicitly we have
\begin{align}
  \vev{V_{\phi,\phi}[\bD^3(x_2)]} = \frac{h_\bD(\phi)}{x_2^3}\,,
  \label{eqn:1ptD}
\end{align}
with the power in the denominator fixed by the conformal dimension of $\bD$ and $h_\bD$ is an 
unknown constant. Likewise,
\begin{align}
  \vev{V_{\phi,\phi}[\bO^2(x_2)]} = \frac{h_\bO(\phi)}{x_2^2}\,.
  \label{eqn:1ptO}
\end{align}

In addition to these bosonic symmetries, when $\phi =\theta$ the crease preserves 
$4 \aQ$'s which offer more relations. Using $\gamma$ for the 6d gamma
matrices and $\rho$ for the R-symmetry matrices (see \cite{Drukker:2020bes} for notations), 
the unbroken supersymmetries are those satisfying
\begin{align}
  (1 + i \rho_1 \gamma_{12}) \aQ_+ = (1 + i \rho_2 \gamma_{13}) \aQ_+ = 0\,.
  \label{eqn:creasesusys}
\end{align}
These are 2 independent constraints, so they select 4
out of the 16 supersymmetries of the $\cN = (2,0)$ theory. Similarly, there are also 
4 preserved $\aS$ conformal supercharges. Note that in this
equation and below we suppress the spinor indices.

Along with the bosonic part, the symmetries assemble into $\osp(4^*|2)$ and
impose further constraints on the correlators~\eqref{eqn:1ptD}
and~\eqref{eqn:1ptO} relating $h_\bD$ and $h_\bO$. To derive them, consider
acting with one of the supersymmetry $\aQ_+$ on the one-point function
\begin{align}
  \vev{\aQ_+ V_{\phi,\phi}[\bQ]} = 0\,.
  \label{eqn:1ptWI}
\end{align}
Away from $x_2 = 0$, the transformations under supersymmetry of $\bQ$ are the
same as on the plane, which were obtained in~\cite{Drukker:2020atp}. They read
(with $a = 1,2$)
\begin{gather}
  \label{eqn:dispsusy}
  \begin{split}
    \aQ \bD_{m} &=
    \frac{1}{2} \gamma_{a m} \partial^a \bQ\,,\\
    \aQ \bQ &=
      2 \gamma_{m} \bD^m
    + 2 \rho_{1i} \gamma_a \partial^a \bO^{i}\,,\\
    \aQ \bO_{i} &=
    \frac{1}{2} \rho_{1i} \bQ\,,
  \end{split}
\end{gather}
for any of the supersymmetries $\aQ$ preserved by the plane.
Restricting these transformations to the subset $\aQ_+$ also preserved by the
crease, we can evaluate~\eqref{eqn:1ptWI} to find
\begin{align}
  \vev{\aQ_+ V_{\phi,\phi}[\bQ]}
  = 2 \left[
    \gamma_3 \vev{V_{\phi,\phi}[\bD^3]}
    - \gamma_{23} \gamma_a \partial^a \vev{V_{\phi,\phi}[\bO^2]}
  \right]
  = 0\,.
\end{align}
To obtain this, we used~\eqref{eqn:creasesusys} to eliminate $\rho$ in favor of
$\gamma$ matrices.  Clearly the derivative is only nonzero along $x_2$. We are left with
\begin{align}
  \vev{V_{\phi,\phi}[\bD^3(x_2)]} = - \partial_2 \vev{V_{\phi,\phi}[\bO^2(x_2)]}
  \qquad \Leftrightarrow \qquad
  h_\bD(\phi) = 2 h_\bO(\phi)\,.
  \label{eqn:nearBPShDhO}
\end{align}

We can now use this constraint to study small deformations of the spherical crease. 
A conformal transformation of the above relations now gives
\begin{align}
  \frac{\vev{V_{\phi,\phi}[\bD^3(u,v)]}}{\vev{V_{\phi,\phi}}} =
  \frac{h_\bD(\phi)}{R^3 \sin^3{u}}\,,
  \qquad
  \frac{\vev{V_{\phi,\phi}[\bO^2(u,v)]}}{\vev{V_{\phi,\phi}}} =
  \frac{h_\bO(\phi)}{R^2 \sin^2{u}}\,.
  \label{eqn:1ptsph}
\end{align}
Following the same logic as in Section~\ref{sec:smallangle}, we can relate the
constants $h_\bD$ and $h_\bO$ to derivatives of the potential.
For instance, integrating the one-point function of $\bO^2$ calculates the small
angle expansion around the BPS case
\begin{align}
  \vev{V_{\phi,\theta}} = \vev{V_{\phi,\phi}} -
   R^2 (\theta-\phi) \int_0^{\pi/2} \cos{u}\, \diff u \int_0^{2\pi} \diff v \vev{V_{\phi,\phi}[\bO^2(u)]} + \dots
\end{align}
Plugging~\eqref{eqn:1ptsph} and integrating we find
\begin{align}
  \frac{1}{\pi} \frac{\diff}{\diff \theta} \exp\left(
  2\pi U(\phi,\theta) \right) \bigg|_{\theta=\phi}
  =
  -2 h_\bO(\phi) \int_0^{\pi/2} \frac{\cos{u} \,\diff u}{\sin^2{u}}
  \int_0^{2\pi} \diff v\,.
\end{align}
The right-hand side contains the integral over $AdS_2$ in the disc topology.
After regularisation this integrates to $-2\pi$. We then obtain
\begin{align}
  h_\bO(\phi) = \frac{1}{4\pi^2} \frac{\diff}{\diff \theta} \exp\left(
  2\pi U(\phi,\theta) \right) \bigg|_{\theta=\phi}\,.
  \label{eqn:hO}
\end{align}
In a similar way $h_\bD$ is related to the derivative with respect to $\phi$.
Integrating the one-point function gives
\begin{align}
  \vev{V_{\phi,\theta}} = \vev{V_{\theta,\theta}} 
+R^2 (\phi-\theta) \int_0^{\pi/2} \cos{u} \,\diff u \int_0^{2\pi} \diff v\,
  e_{\theta_1}^3 \vev{V_\phi[\bD^3(u)]} + \dots
\end{align}
Again the integral gives the volume of $AdS_2$ and leads to
\begin{align}
  h_\bD(\phi) = -\frac{1}{4\pi^2} \frac{\diff}{\diff \phi} \exp\left(
  2\pi U(\phi,\theta) \right) \bigg|_{\theta=\phi}\,.
  \label{eqn:hD}
\end{align}
Finally, comparing~\eqref{eqn:hO} with~\eqref{eqn:hD} and
using~\eqref{eqn:nearBPShDhO} we obtain a relation between the first derivatives
of the potential
\begin{align}
   \frac{\diff U(\phi,\theta)}{\diff \phi} \bigg|_{\phi = \theta} =
   -2\frac{\diff U(\phi,\theta)}{\diff \theta} \bigg|_{\theta = \phi}\,.
\end{align}
Therefore, expanding the potential in Taylor series, we find
\begin{equation}
  \begin{aligned}
  U(\phi,\theta) &=
  U(\phi, \phi) + (\theta - \phi)  \frac{\diff
  U(\phi,\theta)}{\diff \theta} \bigg|_{\theta = \phi} + \dots\\
  &= U(\phi,\phi) + (\phi-\theta) \frac{d U(\phi,\phi)}{\diff \phi} + \dots
  \end{aligned}
\end{equation}
This proves~\eqref{eqn:potentialnearBPS2}.

The explicit expressions in \eqref{eqn:potentialnearBPS} with $C=c$ then gives
\beq
h_\bD=2h_\bO=\frac{c}{2\pi^2}\cos^{2c}\frac{\phi}{2} \tan\frac{\phi}{2}
=-\partial_\phi\left[\frac{1}{4\pi^2}\cos^{2c}\frac{\phi}{2}\right].
\eeq
It would be interesting to try to derive this relation between $h_\bD$ and $c$, 
which would then prove that the leading form of the potential 
is indeed \eqref{eqn:potentialnearBPS}.

\section{Conclusion}
\label{sec:conclusion}

In this paper we defined and studied several realisations of the crease surface
observable in the $\cN = (2,0)$ theory. 
We were able to define a quantity $U(\phi,\theta)$, which can be thought of as a surface-antisurface 
potential, and calculated it in the abelian theory and using classical M2-branes in holography.

For $\theta=\phi$ the surfaces preserve at least eight supercharges and the function 
$U(\phi,\phi)$ takes the simple form \eqref{eqn:potentialBPS} proportional to $\log\cos(\phi/2)$ in both 
calculations. Furthermore, the proportionality constant 
$C$ is equal in both cases to the anomaly coefficient $c$ in \eqref{eqn:anomalycoeffu1}, \eqref{anomalycoef}, 
leading us to conjecture that it is true at any $N$.

It is natural to draw an analogy with cusped Wilson loops in supersymmetric gauge theories 
and in particular $\cN = 4$ SYM. One clear point of deviation are the anomalies---Wilson loop are anomalous 
only in the presence of a cusp, while surface observables are anomalous also for the spherical geometry. 
Furthermore, the natural quantity to associate to the cusp is its anomaly which depends on its angle, while 
the surface anomaly is independent of the angle meaning that the ratio
$\vev{V_{\phi,\phi}}/\vev{V_{S^2}}$ is well-defined and finite.

The comparison is most straightforward then in the BPS case when the cusped Wilson loop is not anomalous, 
but still can provide a finite expectation value, when taking the two longitude configuration on $S^2$ 
(or a compact cusp),
for which the expectation value is related to that of the 
circular Wilson loop $W_\circ(N,\lambda)$~\cite{Drukker:2007dw, Drukker:2007yx, drukker:2007qr, Pestun:2009nn} as
\beq
\vev{W_{\phi,\phi}(N,\lambda)}=\vev{W_{0,0}(N,\tilde\lambda)}\equiv\vev{W_\circ(N,\tilde\lambda)},
\qquad
\tilde\lambda=\left(1-\frac{\phi^2}{\pi^2}\right)\lambda\,,
\eeq
and the expectation value of the circular Wilson loop is
\cite{erickson:2000af,drukker:2000rr,pestun:2007rz}
\beq
\label{circle}
\vev{W_\circ(N,\lambda)}=
\frac{1}{N} L^1_{N-1}\left( -\frac{\lambda}{4N} \right)
e^{\frac{\lambda}{8N}}\,.
\eeq
For the surface operator, we have the potential as calculated from the spherical crease 
in equation~\eqref{eqn:potentialBPS} $U(\phi,\phi)=\frac{c}{\pi}\log\cos(\phi/2)$ with the 
proportionality constant $c$ being the anomaly coefficient \eqref{anomalycoef}.

The BPS Wilson loop depends on three parameters: $N$, $\lambda$ and $\phi$, but the latter 
are combined into the effective coupling $\tilde\lambda$. Still the resulting expression has a rich 
interplay of $N$ and $\tilde\lambda$, while in the absence of a marginal coupling in the 
$\cN = (2,0)$ theory, the BPS potential is a factorised function of the integer $N$ and continuous angle $\phi$. 
For surfaces in higher representations, the expression \eqref{anomalycoef} 
gets modified accordingly \cite{Estes:2018tnu,Chalabi:2020iie,Wang:2020xkc}, 
and the same is true for the Wilson loop \eqref{circle} 
\cite{Drukker:2005kx, Yamaguchi:2006tq, Gomis:2006sb, Hartnoll:2006is, DHoker:2007mci, 
Chen-Lin:2016kkk, Fiol:2013hna, Fiol:2018yuc}.

Although we couldn't prove~\eqref{eqn:potentialBPS} for any $N$,
we did prove rigorously that for small $\phi$ the potential has the form 
\eqref{eqn:potentialnearplane}, with $C=c$, as well as the relation 
\eqref{eqn:potentialnearBPS2} for the near-BPS behaviour.

This is also completely analogous to the cusped Wilson loop. While the functions that appear
are different, in both cases the near-BPS expansion is given by the derivative
with respect to the angle $\phi$ of the BPS result.
For the cusped Wilson loop the generalised 
bremsstrahlung function \cite{correa:2012at} can be written as
\begin{align}
  B(\phi,\lambda,N) =
  -\frac{1}{2}\partial_\phi\log\vev{W_{\phi,\phi}(N,\lambda)}\,,
\end{align}
and the result for the crease~\eqref{eqn:potentialnearBPS}, \eqref{eqn:potentialnearBPS2} is identical.
This relation for the near BPS expansion is guaranteed by the defect CFT analysis in 
Section~\ref{sec:nearBPS}, based on the residual supersymmetry preserved by the crease. 
In the presence of the crease, there is a displacement operator $\bD$ that develops a 1-point function, 
and likewise for its superpartner $\bO$. Their expectation value calculates
first derivatives of the potential, and the relations between them allows to derive~\eqref{eqn:potentialnearBPS2}. 

Going beyond the quadratic order around the plane, or beyond linear deviations from the crease 
requires far more complicated calculations. In the former, it requires the 4-point function of displacements 
and in the latter the 2-point function of the singlet displacement. 
The 4-point function of the displacement operator in the Wilson loop is well 
studied~\cite{Giombi:2017cqn, Liendo:2018ukf, Ferrero:2021bsb, Cavaglia:2021bnz} and nontrivial 
relations among them were found in~\cite{Cavaglia:2022qpg, Drukker:2022pxk}. The same was done 
for the surface operators in \cite{Drukker:2020swu, Drukker:2022pxk}, but it certainly deserves further study.

The potential $U(\phi,\theta)$ is well-defined also away from the BPS limit and we calculated it both 
in the free theory and at large $N$. Here one would wish to invoke the analogy to the generalised cusp 
anomalous dimension of \cite{Drukker:2011za}, with the caveat, that the latter is an anomaly while 
the divergence in the potential is regularised either to zero in the case of the infinite crease, or to a finite 
$-2\pi$ (multiplying $U$) in the case of the spherical crease.

Regardless of this analogy, $U(\phi,\theta)$ allows to interpolate between the plane and the
antiparallel planes (or more properly between the sphere and to coincident hemispheres). 
For planes separated a distance $D$ and with opposite
orientation, it was found in~\cite{maldacena:1997re} that the potential density is
\beq
\mathcal{E} = -\frac{8 N \sqrt{\pi} \,\Gamma(2/3)}{\Gamma(1/6)^3} \frac{1}{D^2}\,.
\label{eqn:potentialantiparallelplanes}
\eeq
The dependence on the distance is fixed by conformal symmetry but the prefactor, which is not,
was one of the first predictions of the $AdS$/CFT correspondence. Among other things, we
generalise this expression to include an R-symmetry angle $\theta$ between the
planes \eqref{eqn:antiparallelpotapprox}. We also obtained the explicit M2-brane for any angle and can calculate
the potential for arbitrary $\phi$ and $\theta$, although the dependence on the angles
is implicit.

Another natural limit is $\phi \to i \infty$, for which the potential grows
linearly in $|\phi|$~\eqref{delta-action}, with no obvious relation
between the prefactors at $N=1$ and $N \to \infty$. This limit can be interpreted as
calculating the action for the lighlike crease in Minkowski space, which we
verify by obtaining the minimal M2-brane in lorentzian $AdS$ and comparing its
action to the potential.  By embedding the
lightlike crease solution in global $AdS$ we also obtain a solution describing two
lightcones glued along a crease, which is the analogous to the 4-cusp
lightlike Wilson loop of~\cite{Kruczenski:2002fb,Alday:2007hr}. It would be interesting 
to see if the lightlike cones and creases have a physical interpretation similar to the 
lightlike Wilson loops, which calculate scattering amplitudes in $\cN=4$ SYM.
The lightlike cusp is also related by an analytic continuation~\cite{Kruczenski:2007cy} to the rotating
string~\cite{Gubser:2002tv}, and it would be interesting to check whether the
solutions presented in Section~\ref{sec:lightlike} are related to the rotating
membrane solutions obtained in~\cite{Sezgin:2002rt,Hartnoll:2002th}.

In this paper we found finite, well-defined observables associated to surface operators in the 
$\cN = (2,0)$ theory, including some BPS protected observables and some that are not. There are 
many more BPS protected observables found in \cite{Drukker:2020bes} that merit further 
study. And many of them may have natural non-BPS extensions. Unraveling them offers a 
window into the dynamics of this mysterious theory.

To point out just one example, there is a family of 1/4 BPS cones interpolating 
between the plane and the cylinder. The BPS cylinder which was studied 
in~\cite{Drukker:2021vyx} has a finite nonzero expectation value. The classical M2-brane solutions 
for the cones of finite angles, whether BPS or not have not been found yet. It should be noted 
that they are related under dimensional reduction to BPS ``latitude''
Wilson loops of 5d SYM~\cite{mezei:2018url} on $S^5$.

Another natural direction is to study creases in surface operators in different theories in varied dimensions.

\section*{Acknowledgements}

We would like to thank Lorenzo Di Pietro for illuminating discussions, and Arkady
Tseytlin, Yifan Wang for their helpful comments on the manuscript.
MT gratefully acknowledges support from King's College London, Universit\'e
Laval,
the Simons Center for Geometry and Physics, Stony Brook University and New York University, 
where part of this project was realised. His visit to KCL was supported by 
ERC Consolidator Grant No. 681908, ``Quantum black holes: A microscopic window into
the microstructure of gravity.''
This research was supported in part by Perimeter Institute for Theoretical Physics. Research at Perimeter
Institute is supported by the Government of Canada through the Department of Innovation, Science and
Economic Development and by the Province of Ontario through the Ministry of Research and Innovation.
ND's research is supported by the Science Technology \& Facilities council under the grants
ST/T000759/1 and ST/P000258/1.

\appendix
\section{Infinite crease in the Poincar\'e patch}
\label{app:poincare}

\subsection{Solution}
\label{app:infinite}
Most of the paper is focused on the compact crease and the holographic section 
(Section~\ref{sec:AdS}) used the $AdS_2\times S^1$ picture. It is interesting to redo the 
calculation for the infinite crease whose holographic description should be in the Poincar\'e 
patch.

As before, we can focus on an $AdS_4\times S^1$ subspace of $AdS_7\times S^4$ and now 
use the metric
\beq
\label{metric-y}
\diff s^2=\frac{y}{L}\left((\diff x^1)^2+\diff r^2+r^2\,\diff\varphi^2\right)+\frac{L^2}{y^2}\left(\diff y^2+y^2\diff\vartheta^2\right).
\eeq
Here $r$ and $\varphi$ are polar coordinates in the $(x^2,x^3)$ plane,  $y$ is the radial coordinate 
in $AdS$, and if we change coordinates to $y=4L^3/z^2$ we get the metric in the form
\beq
\diff s^2=\frac{4L^2}{z^2}(\diff z^2+(\diff x^1)^2+\diff r^2+r^2\,\diff\varphi^2)+L^2\diff\vartheta^2\,.
\eeq

As world-volume coordinates we take $u$, $v$ and $r$ and the ansatz
\beq
z=ur\,,\qquad
x^1=\sqrt{1+u^2}\,v\,,\qquad
\varphi=\varphi(u)\,,\qquad
\vartheta=\vartheta(u)\,.
\eeq
The scaling of $z$ with $r$ is natural because of dilatation invariance of the crease as in the 
case of the cusp \cite{drukker:1999zq}. 
The scaling of 
$x^1$ with $\sqrt{1+u^2}$ may  seem unnatural at first, but it is required to have a full 1d conformal 
symmetry $SO(2,1)$. To see that we write the induced metric
\beq
4L^2\left[\frac{1+u^2}{u^2}\frac{\diff r^2+\diff v^2}{r^2}
+\left(r^2+\frac{u^2v^2}{1+u^2}+r^2\varphi^{\prime2}+\frac{r^2u^2}{4}\vartheta^{\prime2}\right)\frac{\diff u^2}{r^2u^2}
+\frac{2ur\,\diff u\,\diff r+2uv\,\diff u\,\diff v}{r^2u^2}\right].
\eeq
Fixed $u$ slices are clearly $AdS_2$, and hence the required symmetry.

It is possible to solve the equations of motion also without this extra factor of $\sqrt{1+u^2}$ in $x^1$, but the 
solution does not correspond to the conformally invariant crease that we are studying here.

Comparing to \eqref{metric-t}, we see that we should identify $u$ with $t$ and
indeed the resulting lagrangian 
is then identical to \eqref{eqn:creaselagrangian} with the replacement of the measure factors
\beq
\sinh\rho\,\diff\rho\,\diff v\to\frac{\diff r\,\diff v}{r^2}\,.
\eeq
The equations of motion are then the same, but the total action vanishes because
the natural regularisation of the area of $AdS_2$ in this coordinate system is zero. One can 
also implement other regularisations, with a cutoff on $z$ instead of $r$ and possibly restrict 
$x^1$ to a fixed length, rather than set a cutoff on $v$. Those calculations yield different 
(and complicated) expressions.

\subsection{Calibration equations}
\label{app:calibration}

Here we derive the BPS condition~\eqref{BPS} used in section~\ref{sec:BPS}
from the calibration equations obtained in~\cite{Drukker:2020bes}.

Of the four main families of BPS operators identified in~\cite{Drukker:2020bes}, 
the infinite crease falls into two families, type-$\bR$ and
type-$\bH$. For each of the families there exists a calibration form $\phi$ such that 
the surface satisfies the set of equations
\beq
\label{calib}
\partial_m X^M =
\frac{1}{2} g_{ml} \varepsilon^{lnp} G^{ML} \phi_{LNP}
\partial_n X^N \partial_p X^P\,.
\eeq
Here $X^M$ are the target space coordinates with metric $G_{MN}$, $m$ are the 
world-volume indices with induced metric $g_{mn}$ and 
$\varepsilon_{mnp}$ is the Levi-Civita tensor density.

Using the metric
\beq
\diff s^2=\frac{y}{L}\diff x^\mu\diff x^\mu+\frac{L^2}{y^2}\diff y^I\diff y^I\,,
\eeq
with $\mu=1,\dots,6$, $I=1,\dots,5$ and $y=|y|$, 
all surfaces of type $\bR$ are calibrated with respect to the form
\beq
\phi^\bR=-\diff x^1\wedge\sum_{I=1}^5(\diff x^{I+1}\wedge\diff y^I)\,.
\eeq
For the coordinates relevant for the infinite crease~\eqref{metric-y}, this reduces to
\beq
\phi^\bR\big|_{AdS_3\times S^1}
=-\diff x^1\wedge\left[\cos(\varphi-\vartheta)(\diff r\wedge \diff y+ry\,\diff\varphi\wedge\diff\vartheta)
+\sin(\varphi-\vartheta)(y\,\diff r\wedge\diff\vartheta+r\,\diff y\wedge\diff\varphi)
\right].
\eeq

Surfaces of type $\bH$ have the calibration form
\beq
  \phi^\bH =
  \frac{1}{2} \eta_{\mu\nu}^I \diff x^{\mu}\wedge\diff x^\nu\wedge \diff y^I
  -\frac{L^3}{y^3}\diff y_1\wedge\diff y_2\wedge\diff y_3\,,
\eeq
with $\eta^I_{\mu\nu}$ the 't Hooft symbol in 4d
\beq
  \eta^3_{12} = \eta^3_{34} = \eta^2_{31} = \eta^2_{24} = \eta^1_{14} =
  \eta^1_{23} = 1\,.
\eeq
This form is supported on $AdS_5\times S^2$. The crease solution is in $AdS_3\times S^1$, 
but it is useful to keep $AdS_3\times S^2$ components. For the orientation of the 
crease along $x^1$ we should replace 
$y^1\to y^3, y^3\to -y^1$ in the above, and then the calibration form is
\bal
\phi^\bH\big|_{AdS_3\times S^2} 
&=\phi^\bR\big|_{AdS_3\times S^1}
+\diff y^3\wedge\left(\diff x^2 \wedge \diff x^3 - \frac{L^3}{y^3} \diff y^1 \wedge \diff y^2\right)
\\&=
\phi^\bR\big|_{AdS_3\times S^1}
+\diff y^3\wedge\left(r\, \diff r \wedge \diff \varphi - \frac{L^3}{y^2} \diff y \wedge \diff \vartheta\right).
\eal

The extra structure in $\phi^\bH$ is useful because $y^3$ vanishes identically, so taking 
$X^M=y^3$ in~\eqref{calib} we find
\beq
\varepsilon^{lnp} \phi_{y^3NP}
\partial_n X^N \partial_p X^P=0\,,
\eeq
and in particular with the derivatives with respect to $r$ and $u$ (the latter written as prime) this is
\begin{align}
  r \varphi' - \frac{L^3}{y^2} \partial_r(y) \vartheta' =
  r \left( \varphi' + \frac{u^2}{2} \vartheta' \right) = 0\,.
\end{align}
This reproduces the BPS condition~\eqref{BPS} and therefore shows that the
solution with $J_\vartheta = -2 J_\varphi$ is indeed BPS.
Note that the other components of~\eqref{calib} also imply the
rest of the equations of motion~\eqref{eqn:creaseeom2}.

\bibliographystyle{utphys2}
\bibliography{ref}

\end{document}